\documentclass{mn2e}
\usepackage{epsf}
\title{ The CH fraction of Carbon stars at high Galactic latitudes  } 
\title[The CH fraction of Carbon stars at high Galactic latitudes)]{The CH fraction of Carbon stars at high Galactic latitudes}

\author[Aruna Goswami ]{Aruna Goswami$^{1}$\thanks{E-mail:
aruna@iiap.res.in }, Drisya Karinkuzhi$^{1}$, N S Shantikumar$^{2}$ \\ 
$^{1}$Indian Institute of Astrophysics, Koramangala, Bangalore 560034, India\\
$^{2}$Centre for Research and Education in Science and Technology, Indian Institute of Astrophysics, Shidlaghatta Road, Hosakote 562114, India \\
}
\begin{document}

\date{Accepted 2009 October 26.  Received  2009, October 26; in original form 2009 September 17 }


\maketitle

\label{firstpage}

\begin{abstract}
CH stars form a distinct class of objects with characteristic 
properties  like iron deficiency, enrichment of carbon and 
overabundance in heavy
 elements. These properties  can provide strong observational 
constraints for theoretical computation of nucleosynthesis at 
low-metallicity. An important question is the relative surface
density of CH stars  which can provide valuable inputs
to our understanding  on the role of low to intermediate-mass
stars in the early Galactic chemical evolution. Spectroscopic 
characterization provides an effective way of identifying CH stars. 
The present analysis is aimed at a quantitative assessment of 
the fraction of CH stars in a sample of stars  using a set of 
spectral classification 
criteria. The sample consists of ninety two objects selected from
 a collection of candidate  Faint High Latitude Carbon stars from   the
Hamburg/ESO survey. Medium resolution
 (${\lambda}/{\delta\lambda})~ \sim $ 1300 ) spectra  for these 
objects were obtained  using OMR at VBO, Kavalur and HFOSC at HCT, 
IAO, Hanle, during 2007 - 2009  spanning  a wavelength range 
3800 - 6800 \AA\,. Spectral analysis shows 36 of the 92 objects 
to be potential CH stars; combined with our earlier studies 
(Goswami 2005, Goswami et al. 2007) this implies
 ${\sim}$ 37\% ( of 243 ) objects as the CH fraction.
We present  spectral descriptions  of the newly identified CH 
star candidates.  Estimated effective temperatures,
 $^{12}$C/$^{13}$C  isotopic ratios and their locations on the 
two colour J-H vs H-K plot are used to support their 
identification.
\end{abstract}

\begin{keywords} 
stars: CH stars \,-\,variable:
carbon \,-\, stars: spectral characteristics \,-\, stars: AGB \,-\, stars: 
population II 
\end{keywords} 

\section{Introduction}

 Carbon-rich stars ([C/Fe] $\ge$ +1.0) comprise a significant fraction of 
metal-poor ([Fe/H] $\le$ $-$2.0) stars with estimates ranging from 
14 $\pm$ 4\% (Cohen et al. 2005) to 21 $\pm$ 2\% (Lucatello et al. 2005); 
this fraction  increases with decreasing metallicity (Rossi, Beers and Sneden
 1999). A large fraction of carbon-enhanced metal-poor stars exhibit
overabundances of neutron-capture elements relative to iron.  Significant 
insight into the  neutron-capture processes  taking place in the early 
Galaxy  can be derived from  chemical composition studies of  metal-poor 
carbon-stars (Norris et al. 1997a 1997b, 2002; Bonifacio et al. 1998; 
Hill et al. 2000; Aoki et al. 2002a,b;  Goswami et al. 2006, Aoki et al. 2007).
However, formation mechanisms of these stars still remain poorly understood.
 The prime cause of the origin of C-N  stars is believed to be the third 
dredge-up during the AGB evolutionary phase of low to intermediate-mass 
stars; the origin of C-R stars as well as SC-type stars still remains 
unclear (Izzard et al. 2007, Zijlstra  2004). 

 The population II  CH stars,  characterized by a strong G-band of CH and 
s-process elements play significant roles in probing the impact of s-process 
mechanisms in early Galactic chemical evolution. These stars are classified 
into  two distinct types, the Late-type and the Early-type. This classification
 is  based on their  $^{12}$C/$^{13}$C ratios; stars with a large value of
 $^{12}$C/$^{13}$C ratio ( $\ge$ 100) belong to  Late-type, and those
with values of $^{12}$C/$^{13}$C ratio ($\le$ 10) belong to  Early-type.
 The two groups  follow two distinct evolutionary paths. Late-type CH stars 
are further identified as intrinsic stars that  generate s-process elements 
internally and the early-type CH stars as extrinsic stars; they receive 
the s-process  elements via binary mass transfer. The chemical composition 
of early-type  CH stars, if remains unaltered, would   bear signatures 
of the nucleosynthesis  processes operating in the low-metallicity AGB stars.
 Abundance analysis of such stars can provide observational constraints for 
theoretical modelling of $s-$process nucleosynthesis at very low-metallicity
 revealing the time of influence of this process on early Galactic Chemical
 Evolution (GCE). 

Although new large-aperture telescopes has substantially enhanced the number
 of target stars for which high spectral resolution data with high
 signal-to-noise ratio can be obtained,  literature  survey shows  not many 
CH stars have been studied in detail so far. A major difficulty is in 
distinguishing these objects from other types of carbon stars. In  particular,
 Population I C-R, C-N and dwarf carbon stars  exhibit remarkably similar 
spectra with those of carbon giants. It is important to distinguish them 
from one another and  understand  the astrophysical implications  of each 
individual class of stellar population. It is with this motivation  we have
undertaken  to identify the  CH stellar content as well as different types
in a sample of  stars  presented by  Christlieb et al. (2001b). Using low 
resolution spectral analysis we have classified the stars based on a set of
spectral classification criteria. The present  work led 
to the detection of 36 potential CH star candidates among ninety two objects.
Combining this result with  our previous studies we find ${\sim}$ 37\%  
 (of 243) objects are  potential CH star candidates.
This  set of objects  would make important targets for  detailed chemical 
composition  studies  based on high resolution spectroscopy.

Selection of the program stars is outlined in section 2.  Observations 
and data reductions are described in section 3. In section 4 we  briefly 
discuss the main features and spectral characteristics  of C-stars.
 Description of the program stars spectra   and results are drawn in 
section 5.  Conclusions are   presented in section 6. 

\section {Selection of program stars}

The program stars belong to the list of 403  Faint High Latitude Carbon (FHLC) 
stars presented by Christlieb et al. (2001b) from  the database of 
Hamburg/ESO Survey (HES)  described by  Wisotzki et al. (2000).
Hamburg/ESO Survey for carbon stars covers 6400 degree$^{2}$ limited by
${\delta}$ $\le$ +2.5$^{0}$ and $\|$b$\|$ $\ge$ 30$^{0}$. The magnitude limit
is V ${\sim}$ 16.5. The  wavelength range of the spectra is
3200 to 5200 \AA\, 
 at a resolution of 15 \AA\, at H$_{\gamma}$. 
Christlieb et al. found a total of 403 FHLC stars in this survey by
 application of  an 
automated procedure to the digitized spectra. 

The identification of these objects as FHLC stars  was based on a 
measure of line indices - i.e. ratios of the mean photographic densities
 in the carbon molecular absorption features and the continuum band passes.
 The primary consideration is the presence of  strong  C$_{2}$ and CN 
molecular bands shortward of 5200 \AA\,;  CH bands were not considered.

 At high galactic latitudes  different kinds of  carbon stars such as 
N-type carbon stars, dwarf carbon stars, CH-giants, warm  C-R stars
 etc. are known to  populate the region (Green et al. 1994). 
Goswami (2005)
and Goswami et al. (2007) have conducted spectral classification of about 151
objects that belong to the FHLC stars sample offered
 by Christlieb et al. (2001b).  These studies are enhanced by
medium  resolution
spectroscopic analysis  of an additional sample of ninety two objects
 observed during 2007 to 2009.

\section{Observations and Data Reduction}
Observations were  carried out using the 2-m Himalayan Chandra Telescope (HCT)
 at the Indian Astronomical Observatory (IAO), Mt. Saraswati, Digpa-ratsa Ri,
 Hanle.  The spectrograph used is the Himalayan Faint Object Spectrograph
Camera (HFOSC).  HFOSC is an optical imager cum a spectrograph for conducting 
low and medium resolution grism spectroscopy
(http://www.iiap.res.in/centers/iao).
The grism and the camera combination used for observation provided a 
spectral resolution of  $\sim$ 1330( ${\lambda/\delta\lambda}$ );
the observed bandpass ran from about 3800 to 6800 \AA\,. All the objects
listed in Table 1 and 2 are observed during 2007 - 2009.  The B$_{J}$, V, B-V, 
U-B colours listed in the tables are taken from Christlieb et al.(2001b).
Determination of these values  are described in Wisotzki et al. (2000) 
 and  Christlieb et al.(2001a).  B$_{J}$ magnitudes  are accurate to better 
than ${\pm}$0.2mag including zero point errors (Wisotzki et al. 2000).
Spectra of  HD~182040, HD~26, HD~5223, HD~209621, Z~PSc, V460~Cyg and RV~Sct
  used for comparison were obtained  during earlier observations using the 
same observational set up.
A few spectra  acquired using the OMR spectrograph at the cassegrain focus
of the 2.3-m Vainu Bappu Telescope (VBT) at Kavalur, cover a wavelength 
range 4000 - 6100 \AA\, at a resolution of ${\sim}$1000.  With a 600 line
mm$^{-1}$ grating, we get a dispersion of 2.6 \AA\, pixel$^{-1}$.

Observations of  Th-Ar hollow cathode lamp taken immediately before and
after the stellar exposures provide the wavelength calibration. The CCD
data were reduced  using the IRAF software spectroscopic reduction  packages.
For each object two spectra were taken 
and  combined to increase the signal-to-noise ratio.

{\footnotesize
\begin{table*}
\centering
{\bf Table 1: HE stars with prominent C$_{2}$ molecular bands observed during 2007 - 2009 }\\
\tiny
\begin{tabular}{cccccccccccccc}
\hline
Star No.   & RA(2000)$^{a}$ & DEC(2000)$^{a}$& $l$ & $b$ & B$_{J}^{a}$& V$^{a}$ & B-V$^{a}$  &  U-B$^{a}$ &J & H & K &    &Dt of Obs\\
             &            &           &          &          &      &      &       &      &     &    &    &  &    \\
\hline
HE 0008-1712 & 00 11 19.2 & -16 55 34 & 78.5866  & -76.2106 & 16.5 & 15.2 &  1.78 & 1.64 &  13.630 & 13.069 &  12.975 &  & 06.12.2008\\
             &            &           &          &          &      &      &       &      &         &        &         &  &11.09.2008\\
HE 0009-1824 & 00 12 18.5 & -18 07 55 & 75.8376  & -77.2654 & 16.5 & 15.7 &  1.09 & 0.58 &  14.080 & 13.724 &  13.665 &  & 12.09.2008    \\
HE 0037-0654 & 00 40 02.0 & -06 38 13 & 114.8865 & -69.3303 & 16.4 & 15.5 &  1.19 & 0.58 &  14.146 & 13.708 &  13.724 &  & 11.09.2008   \\
HE 0052-0543 & 00 55 00.0 & -05 27 02 & 125.3316 & -68.3057 & 16.5 & 15.0 &  1.95 & 1.74 &  12.952 & 12.241 &  12.086 &  & 12.09.2008    \\
HE 0100-1619 & 01 02 41.6 & -16 03 01 & 136.7651 & -78.6185 & 15.9 & 14.7 &  1.54 & 1.28 &  13.114 & 12.537 &  12.476 &  & 21.11.2008 \\
             &            &           &          &          &      &      &       &      &         &        &         &
&12.09.2008\\
HE 0136-1831 & 01 39 01.8 & -18 16 43 & 176.4932 & -75.9157 & 16.9 & 15.6 &  1.72 & 1.43 &  14.216 & 13.679 &  13.532 &  & 12.09.2008\\
HE 0217+0056 & 02 20 23.5 & +01 10 39 & 163.6094 & -54.5125 & 16.3 & 14.6 &  2.35 & 2.25 &  11.276 & 10.271 &  9.833  &  & 22.11.2008\\
HE 0225-0546 & 02 28 19.4 & -05 32 58 & 174.1738 & -58.4222 & 16.5 & 15.2 &  1.79 & 1.51 &  13.347 & 12.707 &  12.528 &  & 06.12.2008 \\
HE 0228-0256 & 02 31 15.5 & -02 43 07 & 171.5942 & -55.8545 & 16.2 & 14.7 &  1.99 & 1.52 &  12.506 & 11.813 &  11.531 &  & 18.01.2009\\
HE 0308-1612 & 03 10 27.1 & -16 00 41 & 201.1165 & -55.9582 & 12.5 &  --  &  --   &  --  &  10.027 & 9.475  &  9.331  &  & 21.11.2008\\
HE 0420-1037 & 04 22 47.0 & -10 30 26 & 205.0454 & -37.7176 & 15.2 & 14.7 &  1.38 & 0.99 &  12.341 & 11.815 &  11.695 &  & 21.11.2008  \\
HE 0945-0813 & 09 48 18.7 & -08 27 40 & 245.0790 & 33.1497 & 16.2 & 15.3  & 1.22 & 1.11  & 13.531 &  13.026  &     12.903  &  &  08.04.2007\\
             &            &           &          &          &      &      &       &      &        &          &         & &09.04.2007\\
HE 0954+0137 & 09 57 19.2 & +01 23 00 & 237.1030 &  41.0018 & 16.6 & 15.7 & 1.18  & 0.32 &   --    &   --   &  --     &  & 21.11.2008\\
HE 1011-0942 & 10 14 25.0 & -09 57 54 & 251.7699 &  36.8590 & 15.4 & 14.2 & 1.65  & 1.76 &  11.209 & 10.432 &  10.125 &  & 22.11.2008\\
HE 1019-1136 & 10 22 14.7 & -11 51 39 & 255.1421 &  36.8087 & 15.2 & 13.9 & 1.84  & 2.29 &  10.005 & 9.031  &  8.488  &  & 06.02.2009  \\  
HE 1027-2501 & 10 29 29.5 & -25 17 16 & 266.6832 &  27.4163 & 13.9 & 12.7 & 1.73  & 1.51 &  10.627 & 9.896 & 9.722    &  & 22.11.2008\\
HE 1045-1434 & 10 47 44.1 & -14 50 23 & 263.5905 &  38.4049 & 15.5 & 14.6 &  1.23 & 0.96 &  12.935 & 12.449 & 12.244 &  &  09.04.2007\\
HE 1051-0112 & 10 53 58.8 & -01 28 15 & 253.5294 &  49.7900 & 17.0 & 16.0 & 1.44  & 0.94 &  14.347 & 13.794 &  13.703 &  & 06.12.2008 \\
HE 1102-2142 & 11 04 31.2 & -21 58 29 & 272.5241 &  34.5071 & 16.0 & 14.9 & 1.44 & 0.98  &  13.275 & 12.714 &  12.601 &  & 16.03.2009  \\
HE 1110-0153 & 11 13 02.7 & -02 09 28 & 261.1493 &  52.4883 & 16.5 & 15.5 & 1.47 & 1.29  &  12.912 & 12.205 &  12.063 &  & 04.04.2009 \\
HE 1116-1628 & 11 19 03.9 & -16 44 50 & 273.2181 &  40.7347 & 16.6 & 15.6 & 1.28 & 1.41  &  13.241 & 12.614 & 12.482  &  & 06.02.2009 \\
HE 1119-1933 & 11 21 43.5 & -19 49 47 & 275.7342 &  38.2583 & 12.8 & 14.6 & 1.34 & 0.87  &  13.043 & 12.571 & 12.422  &  & 18.03.2009  \\
HE 1120-2122 & 11 23 18.6 & -21 38 33 & 277.1276 &  36.7743 & 12.9 &  --  &  --  &  --   &  9.573  & 8.902  & 8.788   &  & 17.03.2009 \\
HE 1123-2031 & 11 26 08.7 & -20 48 19 & 277.4490 &  37.8097 & 16.8 & 15.8 & 1.33 & 1.19  &  13.513 & 12.940 & 12.800  &  & 18.03.2009\\
HE 1127-0604 & 11 29 36.8 & -06 20 51 & 269.3427 &  51.1090 & 16.8 & 15.9 & 1.23 & 0.87  &  14.594 & 14.087 & 14.027  &  & 18.01.2009 \\
HE 1142-2601 & 11 44 52.9 & -26 18 29 & 284.8485 &  34.2157 & 13.9 & 13.0 & 1.28 & 1.07  &  11.218 & 10.675 & 10.539  &  
& 04.04.2009\\
HE 1145-1319 & 11 48 21.4 & -13 36 38 & 280.3756 &  46.4800 & 16.4 & 15.4 & 1.37 & 1.32  &  13.466 & 12.934 & 12.790  &  & 17.03.2009\\
             &            &           &          &          &      &      &       &      &         &        &         &   &05.04.2009\\
HE 1146-0151 & 11 49 02.3 & -02 08 11 & 273.2807 & 57.0990 & 14.9  & 14.2 & 0.96 & 0.97  &  12.929 & 12.400 & 12.262  &  & 16.03.2009\\
HE 1157-1434 &  12 00 11.5 &  -14 50 50 & 284.8854  &   46.2211   &    16.1 &  13.8  &  1.62  &  1.41& 11.792   &  11.178  &     11.019    &   &  06.06.2007\\
             &            &           &          &          &      &      &       &      &        &          &         & &07.06.2007\\
HE 1205-0521 & 12 07 53.  & -05 37 51 & 283.5026 & 55.5893 & 15.2 & 14.4 & 1.09  &0.48   & 12.434  & 11.988 &  11.806 &  & 04.04.2009\\
HE 1205-2539 & 12 08 08.1 & -25 56 38 & 290.8957 & 35.9137 & 13.6 & 15.4 & 1.91  &1.72   & 12.313  & 12.665  & 12.540 &  & 17.03.2009 \\
HE 1210-2636 & 12 12 59.7 & -26 53 22 & 292.4334 & 35.1974 & 13.8 & 12.6 & 1.65  &1.52   & --      &   --   &   --    &  & 06.05.2009\\
HE 1228-0417 & 12 31 12.5 & -04 33 40 & 293.4045 & 57.9360 & 14.8 & 14.0 & 1.17  &0.96   &  12.406 &  11.969 & 11.818 &  & 04.04.2009\\
HE 1230-0327 & 12 33 24.1 & -03 44 28 & 294.2159 & 58.8251 & 13.7 & 15.1 & 1.02  &0.80   &  13.683 &  13.224 & 13.100 &  & 17.03.2009\\
HE 1238-0836 &  12 41 02.4 &  -08 53 06 & 298.5709  &   53.8987   &    13.1 &  --      &   --      & --     &    9.189  &   8.503   & 8.154  &    & 08.04.2007\\
HE 1253-1859 & 12 56 38.4 & -19 15 32 & 304.6277 & 43.5957 & 13.7 & 12.9 & 1.09  &1.22   &  10.604 &  9.972  & 9.850  &  & 16.03.2009\\
HE 1315-2035 & 13 17 57.4 & -20 50 53 & 311.2275 & 41.5961 & 16.7 & 15.7 & 1.35  &0.93   &  13.678 &  13.236 & 13.047 &  & 06.05.2009\\
HE 1331-2558 & 13 34 20.1 & -26 13 38 & 314.7873 & 35.6550 & 13.9 & 16.0 & 1.53  &1.10   &  10.998 &  10.394 & 10.240 &  & 18.03.2009\\
HE 1344-0411 & 13 47 25.7 & -04 26 04 & 328.2429 & 55.6614 & 16.3 & 14.9 & 2.02  &2.29   &  11.708 &  10.848 & 10.467 &  & 05.04.2009\\
HE 1358-2508 & 14 01 12.3 & -25 22 39 & 322.8426 & 34.4322 & 13.2 & 12.3 & 1.25  &0.84   & 9.521  &   8.802  & 8.532  &  & 18.01.2009\\
HE 1400-1113 & 14 03 39.8 & -11 28 04 & 329.7145 & 47.6129 & 16.0 & 15.2 & 1.19  &0.75   & 13.838  &  13.411 & 13.276 & 
& 17.03.2009\\
HE 1404-0846 &  14 06 55.1 &  -09 00 58 & 332.3876  &   49.4897   &   15.3 &  14.2  &  1.52   & 1.29  &  12.435   &11.783& 11.681&   &  07.06.2007\\             
HE 1405-0346 & 14 07 58.3 & -04 01 03 & 336.5341 & 53.7853 & 14.7 & 13.5 & 1.68  &1.28   & 11.608  &  11.068 & 10.903 & 
& 16.03.2009\\
HE 1410-0125 & 14 13 24.7 & -01 39 54 & 340.6313 & 55.0908 &  --  &  --  &   --  &  --   & 10.360  &  9.770  & 9.651   & & 16.03.2009\\
             &            &           &          &          &      &      &       &      &         &         &         & &11.06.2008\\
HE 1418-0306 &14 20 57.1 & -03 19 54 & 341.7366  & 52.6618  &  --  & 13.0 &  1.66 & 1.11 &  10.505 & 9.767   &  9.505 &   &  09.04.2007\\   
HE 1425-2052 & 14 28 39.5 & -21 06 05 & 331.4023 & 36.3382 & 13.6 & 12.7 & 1.27  &1.29   & 10.043  &  9.446   & 9.273  & & 17.03.2009\\
HE 1428-1950 & 14 30 59.4 & -20 03 42 & 332.6046 & 37.0083 &  --  &  --  &  --    &  --  & 9.988   &  9.469  & 9.318   & & 06.02.2009\\
HE 1429-1411 & 14 32 40.6 & -14 25 06 & 336.6322 & 41.7341 & 12.5 & 11.1  &1.99  &1.89   & 7.622   &  6.721  & 6.346   & & 06.02.2009\\
HE 1430-0919 & 14 33 12.9 & -09 32 53 & 340.3560 & 45.7983 & 14.9 & 13.8  &1.46  &0.75   & 12.476  &  11.971 &  11.862 & & 18.03.2009 \\
             &            &           &          &          &      &      &       &      &         &         &         & &09.05.2007\\
HE 1431-0755 &  14 34 32.7 &  -08 08 37 & 341.8828  &   46.7820     & 14.6  & 13.5 &   1.51 &   1.44  &   11.283  & 10.605    &  10.422  &   &  08.05.2007\\ 
HE 1439-1338 & 14 42 26.4 & -13 51 18 & 339.7039 & 40.9588 & 14.5 & 13.5  &1.39  &0.97   & 10.810  &  10.112 &   9.836 & & 25.07.2008\\
             &            &           &          &          &      &      &       &      &         &         &         & &09.04.2007\\
HE~1440-1511 &  14 43 07.1  & -15 23 48 & 338.737   &   39.5765   & 14.6  & 13.8   & 1.13   & 0.87 &  12.238  & 11.757 & 11.606    &    &  09.05.2007  \\
HE 1442-0058 &  14 44 48.9 &  -01 10 56 & 351.4558  &   50.6883    & 17.8  & 16.2  &  2.15  &  1.80 &  12.287   &   11.268    &   10.751   &   & 06.06.2007\\ 
HE 1447+0102 & 14 50 15.1 & +00 50 15 & 355.2263 & 51.2262 & 15.6 & 15.0  &0.90  &0.14   & 13.207  &  12.760 &  12.682 & & 12.09.2008\\
             &            &           &          &          &      &      &       &      &         &         &         & &08.04.2007\\
HE 1525-0516 & 15 27 52.2 & -05 27 04 & 358.1110 & 40.1019 & 16.8 & 15.8  &1.29  &1.14   & 13.972  &  13.479 &  13.314 & & 11.09.2008 \\
             &            &           &          &          &      &      &       &      &        &          &         & &07.06.2007\\
             &            &           &          &          &      &      &       &      &        &          &         & &08.06.2007\\
HE 2114-0603 & 21 17 20.8 & -05 50 48 & 45.5467  & -34.9459& 16.7 & 15.4  &1.80  &1.58   & 12.472  &  11.786 &  11.615 & & 11.09.2008 \\
             &            &           &          &          &      &      &       &      &        &          &         & &09.05.2007\\
HE 2144-1832 &  21 46 54.7 &  -18 18 15 & 34.6476 & -46.7834& 12.6 &   -- &   --  &   -- & --     & --       &  --     &   & 06.06.2007\\
HE 2157-2125 &  22 00 25.5 &  -21 11 23 & 32.0715 &-50.7274 & 16.8 & 15.9 & 1.29  & 1.02 &  14.389&  13.980  & 13.742  &    & 08.06.2007\\
             &            &           &          &          &      &      &       &      &        &          &         & &12.09.2008\\
HE 2211-0605 & 22 13 53.5 & -05 51 06 & 55.3066  & -46.9505& 16.0 & 15.1  &1.21  &1.01   & 13.383 &  12.875  &  12.727 & & 24.07.2008 \\
HE 2213-0017 & 22 15 37.1 & -00 02 59 & 62.3019  & -43.8276& 16.4 & 14.6  &2.38  &2.52   & 10.860 &   9.832  &   9.351 & & 24.07.2008\\
             &            &           &          &          &      &      &       &      &        &          &         & &27.05.2007\\
HE 2216-0202 & 22 18 47.5 & -01 47 36 & 61.0859  & -45.5370& 17.2 & 16.4  &1.01  &0.43   & 14.571 &  14.165  &  14.028 & & 06.12.2008\\
             &            &           &          &          &      &      &       &      &        &          &         & &11.09.1008\\
             &            &           &          &          &      &      &       &      &        &          &         & &24.07.2009\\
HE 2222-2337 & 22 25 38.8 & -23 22 44 & 31.2015  & -56.9371& 17.1 & 15.7  &1.93  &1.84   & 13.535 &  12.837  &  12.664 & & 12.09.2008\\
HE 2225-1401 & 22 28 10.7 & -13 46 23 & 47.4570  & -54.0483& 16.5 & 14.5  &2.72  &2.36   & 11.870 &  10.748  &  9.896  & & 21.11.2008\\
HE 2228-0137 & 22 31 26.2 & -01 21 42 & 64.4957  & -47.7030& 15.8 & 14.7  &1.56  &1.20   & 12.301 &  11.715  &  11.589 & & 11.09.2008\\
             &            &           &          &          &      &      &       &      &        &          &         & &25.07.2009\\
HE 2246-1312 & 22 49 26.4 & -12 56 35 & 53.2930  & -58.1569& 17.0 & 15.9  &1.56  &1.60   & 14.101 &  13.472  &  13.303 & & 11.09.2008\\
             &            &           &          &          &      &      &       &      &        &          &         & &25.07.2008\\
HE 2255-1724 & 22 58 06.8 & -17 08 19 & 47.9045  & -62.0030& 16.1 &15.1   &1.45  &1.05   & 12.931 &  12.374  &  12.310 & & 11.09.2008\\
             &            &           &          &          &      &      &       &      &        &          &         & &25.07.2008\\
HE 2305-1427 & 23 08 10.9 & -14 11 27 & 55.9986  & -62.6875& 16.4 &15.3   &1.41  &0.80   & 13.269 &  12.762  &  12.744 & & 25.07.2008\\
             &            &           &          &          &      &      &       &      &        &          &         & &10.10.2008\\
HE 2334-1723 & 23 37 03.7 & -17 06 34 & 59.3241  & -70.1108& 16.1  &  15.1 &1.38   &1.48 & 13.583 &  13.119  &  12.977 & & 10.10.2008\\
             &            &           &          &          &      &      &       &      &        &          &         & &25.07.2008\\
HE 2347-0658 & 23 49 56.3 & -06 41 55 & 84.5844  & -64.8865& 16.9 &16.0   &1.30  & 0.69  & 14.568 &  14.145  &  14.158 & & 11.09.2008 \\
             &            &           &          &          &      &      &       &      &        &          &         & &24.07.2008\\
HE 2353-2314 & 23 55 44.0 & -22 58 09 & 48.1281  & -76.7257& 16.5 &15.2   &1.82  & 1.53  & 13.064 &  12.403  &  12.199 & & 25.07.2008  \\
\hline
\end{tabular}

$^{a}$ From Christlieb et al. (2001b)\\

\end{table*}
}

{\footnotesize
\begin{table*}
\centering
{\bf Table 2:  HE stars without prominent C$_{2}$ molecular bands observed during 2007 - 2009 }\\
\tiny
\begin{tabular}{cccccccccccccc}
\hline
Star No.   & RA(2000)$^{a}$ & DEC(2000)$^{a}$& $l$ & $b$ & B$_{J}^{a}$& V$^{a}$ & B-V$^{a}$  &  U-B$^{a}$ &J & H & K &   mol. bands &Dt of Obs\\
             &            &           &          &          &      &      &       &      &     &    &    &  &    \\
\hline
HE 0422-2518 & 04 24 38.5 & -25 12 10 & 223.2761 & -42.4854 & 13.9 &  --  &  --   & --   & --      &  --    &   --    & CH, CN & 22.11.2008\\
HE 0443-2523 & 04 45 17.7 & -25 17 48 & 224.9946 & -38.0024 & 13.8 & --   &    -- &  --  &  10.704 & 10.169 &  10.012 &  CH, CN & 06.12.2008 \\
HE 0513-2008 & 05 15 14.9 & -20 04 59 & 221.6222 & -29.8139 & 13.2 & --   &  --   &  --  &  9.398  &  8.810 &  8.677  &  CH, CN& 18.01.2009\\
             &            &           &          &          &      &      &       &      &         &        &         &  &22.11.2008\\
HE 1027-2644 & 10 29 57.8 & -26 59 51 & 267.8862 &  26.0820 & 14.4 & 13.4 & 1.40  & 1.59 &  --     &  --    &   --    &        & 17.01.2009\\
HE 1033-0059 & 10 36 34.0 & +00 43 39 & 246.4726 &   48.2458& 12.9 & 13.0 & 1.10  &  1.29 & 10.709 & 10.138 & 9.951   &  CH, CN & 09.04.2007\\
HE 1036-2615 & 10 38 25.9 & -26 30 50 & 269.3336 &  27.5445 & 14.6 & 13.7 & 1.21  & 1.49 & --      &  --    &  --     &  CH, CN& 17.01.2009\\
HE 1037-2644 & 10 40 02.3 & -27 00 36 & 269.9776 &  27.3253 & 14.3 & 13.4 & 1.33  & 1.38 & 11.360  & 10.776 & 10.662  &  CH, CN& 06.02.2009 \\
HE 1056-1855 & 10 59 12.2 & -19 11 08 & 269.4843 &  36.2931 & 13.6 &  --  &  --   &  --  &  10.784 & 10.249 &  10.090 &  CH, CN& 22.11.2008  \\
HE 1104-1442 & 11 06 30.3 & -14 58 56 & 268.5996 &  40.7912 &  --  & 13.7 &  1.31 & 1.16 & 12.217  & 11.568 &  11.470 & 
CH, CN  &  09.04.2007\\
HE 1105-2736 & 11 07 44.7 & -27 52 35 & 276.5372 &  29.6271 & 14.1 & 13.2 & 1.30 & 1.23  &11.314   & 10.742 & 10.551  &  CH, CN& 17.03.2009\\
HE 1112-2557 & 11 15 14.1 & -26 13 28 & 277.4416 &  31.8419 & 14.5 & 13.6 & 1.29 & 1.45  & 11.548  & 11.006 & 10.849  &  CH, CN& 06.05.2009\\
HE 1150-2546 & 11 53 15.5 & -26 03 41 & 286.9462 & 34.9979 & 15.6  &  --  &  --  &   --  &   --     &  --   &  --     &  & 18.03.2009\\
HE 1152-2432 & 11 54 35.0 & -24 48 44 & 286.8824 & 36.2813 &  --   &  --  &  --  &  --  &   9.413 & 8.833  &  8.686   &  CH, CN& 18.03.2009 \\
HE 1229-1857 & 12 31 46.8 & -19 14 02 & 296.5412 & 43.3938 & 14.6  & 14.0 & 0.85  &0.73   &  12.124 &  11.569 & 11.459 &  CH, CN& 04.04.2009\\
HE 1236-0036 & 12 39 25.4 & -00 52 58 & 296.5586 & 61.8402 &  16.5 &  13.5 & 0.92 & 0.94  &11.679   &  11.176 & 11.104 & CH, CN  &  08.05.2007\\ 
HE 1406-2016 & 14 09 44.1 & -20 30 57 & 326.6476 & 38.7233 & 15.1 & 14.2 & 1.28  &1.22   & 12.092  &  11.558 &  11.423 & CH, CN& 16.03.2009 \\
HE 1420-1659 & 14 23 03.1 & -17 12 51 &332.7845  &39.8992  & 13.1   &  --  &  --   &  --   & 9.871   &  9.226  & 9.067 &
CH, CN & 18.01.2009 \\
HE 1514-0207 & 15 16 38.9 & -02 18 33 & 358.6655 & 44.2895 & 13.6 &  --   &  --  &  --   &  10.335 &  9.703  &  9.536  & CH, CN& 24.07.2008\\
HE 2115-0709 & 21 17 42.1 & -06 57 11 & 44.4121  &-35.5571 & 15.1 & 16.2  & 1.68 &  1.70 &  12.673 &  12.000 &  11.808 & CH, CN &  07.06.2007\\
HE 2121-0313 & 21 23 46.2 & -03 00 51 & 49.5147  & -34.9016& 14.9 & 13.9  &1.35  &1.47   & 11.304  &  10.718 &  10.592 & CH, CN& 11.06.2008 \\
HE 2124-0408 & 21 27 06.8 & -03 55 22 & 49.0907  & -36.0859& 14.8 & 13.9  &1.26  &1.15   & 11.578  &  10.986 &  10.848 & CH, CN& 24.07.2008  \\
HE 2205-1033 & 22 08 29.2 & -10 18 37 & 48.6453  & -48.1654& 12.8 &  --   &  --  &  --   & 9.732  &  9.178   &  8.980  & CH, CN& 24.07.2008\\
\hline
\end{tabular}
$^{a}$ From Christlieb et al. (2001b)\\

\end{table*}
}

{\footnotesize
\begin{table*}
\centering
{\bf Table 3: Potential CH star candidates}\\
\tiny
\begin{tabular}{cccccccccccccc}
\hline
Star No.   & RA(2000)$^{a}$ & DEC(2000)$^{a}$& $l$ & $b$ & B$_{J}^{a}$& V$^{a}$ & B-V$^{a}$  &  U-B$^{a}$ &J & H & K & &Dt of Obs\\
             &            &           &          &          &      &      &       &      &     &    &    &  &    \\
\hline
HE 0008-1712 & 00 11 19.2 & -16 55 34 & 78.5866  & -76.2106 & 16.5 & 15.2 &  1.78 & 1.64 &  13.630 & 13.069 &  12.975 &  & 06.12.2008\\
             &            &           &          &          &      &      &       &      &         &        &         &  &11.09.2008\\
HE 0052-0543 & 00 55 00.0 & -05 27 02 & 125.3316 & -68.3057 & 16.5 & 15.0 &  1.95 & 1.74 &  12.952 & 12.241 &  12.086 &  & 12.09.2008    \\
HE 0100-1619 & 01 02 41.6 & -16 03 01 & 136.7651 & -78.6185 & 15.9 & 14.7 &  1.54 & 1.28 &  13.114 & 12.537 &  12.476 &  & 21.11.2008 \\
             &            &           &          &          &      &      &       &      &         &        &         &
&12.09.2008\\
HE 0136-1831 & 01 39 01.8 & -18 16 43 & 176.4932 & -75.9157 & 16.9 & 15.6 &  1.72 & 1.43 &  14.216 & 13.679 &  13.532 &  & 12.09.2008\\
HE 0225-0546 & 02 28 19.4 & -05 32 58 & 174.1738 & -58.4222 & 16.5 & 15.2 &  1.79 & 1.51 &  13.347 & 12.707 &  12.528 &  & 06.12.2008 \\
HE 0308-1612 & 03 10 27.1 & -16 00 41 & 201.1165 & -55.9582 & 12.5 & --   &  --   &  --  &  10.731 & 9.821  &  9.406  &  & 21.11.2008\\
HE 0420-1037 & 04 22 47.0 & -10 30 26 & 205.0454 & -37.7176 & 15.2 & 14.7 &  1.38 & 0.99 &  12.341 & 11.815 &  11.695 &  & 21.11.2008  \\
HE 1027-2501 & 10 29 29.5 & -25 17 16 & 266.6832 &  27.4163 & 13.9 & 12.7 & 1.73  & 1.51 & 10.627 & 9.896    & 9.722  &  & 22.11.2008\\
HE 1045-1434 & 10 47 44.1 & -14 50 23 & 263.5905 &  38.4049 & 15.5 & 14.6 &  1.23 & 0.96 &  12.935 & 12.449 & 12.244 &  &  09.04.2007\\
HE 1051-0112 & 10 53 58.8 & -01 28 15 & 253.5294 &  49.7900 & 17.0 & 16.0 & 1.44  & 0.94 &  14.347 & 13.794 &  13.703 &  & 06.12.2008 \\
HE 1102-2142 & 11 04 31.2 & -21 58 29 & 272.5241 &  34.5071 & 16.0 & 14.9 & 1.44  & 0.98 &  13.275 & 12.714 &  12.601 &  & 16.03.2009  \\
HE 1110-0153 & 11 13 02.7 & -02 09 28 & 261.1493 &  52.4883 & 16.5 & 15.5 & 1.47 & 1.29  &  12.912 & 12.205 &  12.063 &  & 04.04.2009 \\
HE 1119-1933 & 11 21 43.5 & -19 49 47 & 275.7342 &  38.2583 & 12.8 & 14.6 & 1.34 & 0.87  &  13.043 & 12.571 & 12.422  &  & 18.03.2009  \\
HE 1120-2122 & 11 23 18.6 & -21 38 33 & 277.1276 &  36.7743 & 12.9 &  --  &  --  &  --   &  9.573  & 8.902  & 8.788   &  & 17.03.2009 \\
HE 1123-2031 & 11 26 08.7 & -20 48 19 & 277.4490 &  37.8097 & 16.8 & 15.8 & 1.33 & 1.19  &  13.513 & 12.940 & 12.800  &  & 18.03.2009\\
HE 1142-2601 & 11 44 52.9 & -26 18 29 & 284.8485 &  34.2157 & 13.9 & 13.0 & 1.28 & 1.07  & 11.218  & 10.675 & 10.539  &  
& 04.04.2009\\
HE 1145-1319 & 11 48 21.4 & -13 36 38 & 280.3756 &  46.4800 & 16.4 & 15.4 & 1.37 & 1.32  &  13.466 & 12.934 & 12.790  &  & 17.03.2009\\
             &            &           &          &          &      &      &       &      &         &        &         &
&05.04.2009\\
HE 1146-0151 & 11 49 02.3 & -02 08 11 & 273.2807 & 57.0990 & 14.9  & 14.2 & 0.96 & 0.97  &  12.929 & 12.400 & 12.262  &  & 16.03.2009\\
HE 1157-1434 & 12 00 11.5 & -14 50 50 & 284.8854 & 46.2211 & 16.1  & 13.8 & 1.62 &  1.41 &  11.792 & 11.178 & 11.019  &  & 06.06.2007\\
             &            &           &          &         &       &      &      &       &         &        &         &  &07.06.2007\\
HE 1205-2539 & 12 08 08.1 & -25 56 38 & 290.8957 & 35.9137 & 13.6 & 15.4 & 1.91  &1.72   & 13.313 & 12.665  & 12.540  &  & 17.03.2009 \\
HE 1210-2636 & 12 12 59.7 & -26 53 22 & 292.4334 & 35.1974 & 13.8 & 12.6 & 1.65  &1.52   &   --   &  --     &   --    &  & 06.05.2009\\
HE 1228-0417 & 12 31 12.5 & -04 33 40 & 293.4045 & 57.9360 & 14.8 & 14.0 & 1.17  &0.96   &  12.406 &  11.969 & 11.818 &  & 04.04.2009\\
HE 1253-1859 & 12 56 38.4 & -19 15 32 & 304.6277 & 43.5957 & 13.7 & 12.9 & 1.09  &1.22   &  10.60  &  9.972  & 9.850  &  & 16.03.2009\\
HE 1331-2558 & 13 34 20.1 & -26 13 38 & 314.7873 & 35.6550 & 13.9 & 16.0 & 1.53  &1.10   &  10.998 &  10.394 & 10.240 &  & 18.03.2009\\
& 17.03.2009\\
HE 1404-0846 & 14 06 55.1 & -09 00 58 & 332.3876 & 49.4897 & 15.3 & 14.2 & 1.52  &1.29   &  12.435 &11.783   & 11.681&  & 07.06.2007\\             
HE 1405-0346 & 14 07 58.3 & -04 01 03 & 336.5341 & 53.7853 & 14.7 & 13.5 & 1.68  &1.28   & 11.608  &  11.068 & 10.903 &  & 16.03.2009\\
HE 1410-0125 & 14 13 24.7 & -01 39 54 & 340.6313 & 55.0908 & --   &  --  &  --   &  --   & 10.360  &  9.770  & 9.651   & & 16.03.2009\\
             &            &           &          &          &      &      &       &      &         &         &         & &11.06.2008\\
HE 1425-2052 & 14 28 39.5 & -21 06 05 & 331.4023 & 36.3382  & 13.6 & 12.7 & 1.27  &1.29   & 10.043 & 9.446   & 9.273   & & 17.03.2009\\
HE 1431-0755 & 14 34 32.7 & -08 08 37 & 341.8828 & 46.7820  & 14.6 & 13.5 & 1.51  & 1.44  &  11.283  & 10.605 &  10.422 & & 08.05.2007\\ 
HE~1440-1511 & 14 43 07.1 & -15 23 48 & 338.737  & 39.5765  & 14.5 & 13.5 & 1.13  & 0.87  &  12.238  & 11.757 & 11.606 &  &  09.05.2007  \\
HE 1447+0102 & 14 50 15.1 & +00 50 15 & 355.2263 & 51.2262  & 15.6 & 15.0 &  0.90 & 0.14  & 13.207   & 12.760 & 12.682 &  &   08.04.2007\\
HE 1525-0516 & 15 27 52.2 & -05 27 04 & 358.1110 & 40.1019 & 16.8 & 15.8  &1.29  &1.14   & 13.972  &  13.479 &  13.314 & & 11.06.2008 \\
             &            &           &          &          &      &      &       &      &         &         &         & &07.06.2007\\
             &            &           &          &          &      &      &       &      &         &         &         & &08.06.2007\\
HE 2114-0603 & 21 17 20.8 & -05 50 48 & 45.5467  & -34.9459& 16.7 & 15.4  &1.80  &1.58   & 12.472  &  11.786 &  11.615 & & 11.09.2008 \\
             &            &           &          &          &      &      &       &      &         &         &         & &09.05.2007\\
HE 2211-0605 & 22 13 53.5 & -05 51 06 & 55.3066  & -46.9505& 16.0 & 15.1  &1.21  &1.01   & 13.383 &  12.875  &  12.727 & & 24.07.2008 \\
HE 2228-0137 & 22 31 26.2 & -01 21 42 & 64.4957  & -47.7030& 15.8 & 14.7  &1.56  &1.20   & 12.301 &  11.715  &  11.589 & & 11.09.2008\\
             &            &           &          &          &      &      &       &      &        &          &         & &25.07.2009\\
HE 2246-1312 & 22 49 26.4 & -12 56 35 & 53.2930  & -58.1569& 17.0 & 15.9  &1.56  &1.60   & 14.101 &  13.472  &  13.303 & & 11.09.2008\\
             &            &           &          &          &      &      &       &      &        &          &         & &25.07.2008\\
\hline
\end{tabular}

$^{a}$ From Christlieb et al. (2001b)\\

\end{table*}
}

\section { Characteristics of carbon stars and spectral classification}

 Spectral classification  helps reducing
 the number of stars to  be analyzed to a tractable number of
prototype objects of different groups;  each group may be   correlated
with one or more physical parameters such as luminosity and temperature.
Abundance anomalies  observed in carbon stars  cannot be explained on 
the  basis of observed temperature and luminosity of the stars; 
 it is therefore  difficult to 
devise a classification scheme for carbon stars  based  on  only 
these two physical parameters. 
Morgan-Keenan system  for carbon star classification (Keenan 1993)
divided  carbon stars into C-R, C-N and C-H sequence,
 with subclasses running to C-R6, C-N6 and C-H6
according to temperature criteria. In the old R-N system, CH stars
that were classified as R-peculiar are put in a separate class in the
new system. 
In the following we  briefly  discuss  the main characteristics that place 
carbon stars into different groups. Detailed discussions on  carbon stars are
available in literature  including Wallerstein and Knapp  (1998) and 
references therein.

In contemporary  stellar  classification schemes  assigning stars to
 `morphological groups' is  largely in practice. 
Carbon stars are  primarily  classified  based on the strength of carbon
molecular bands.  
 The C-N stars are characterized by strong depression 
of light in the violet part of the spectrum. The cause of rapidly weakening 
continuum below about 4500\AA\, is not fully established yet, but believed
to be due to scattering by particulate matter. Oxygen-rich stars of similar 
effective temperatures do not show such weakening.
 C-N stars  are also  easily detected  for  their characteristic 
infrared colours. The majority of C-N stars show ratios of $^{12}$C/$^{13}$C 
in the range of  30 to 100 while in C-R stars this ratio ranges from
 4 to 9 (Lambert et al. 1986). They have  lower temperatures and stronger 
molecular bands than those of C-R stars. They are used as tracers of an 
intermediate age population in extragalactic objects.  

CH stars are  characterised by the strong G-band of CH in their spectra. They
 form a group of warm stars of equivalent spectral
types, G and K normal giants, but  show  weaker metallic lines.
In general, CH stars are high velocity objects, large radial velocities
indicating that  they belong to the halo population of the Galaxy 
(McClure 1983, 1984, McClure \& Woodsworth 1990).
 `CH-like' stars, 
where CH are less dominant have low space velocities (Yamashita 1975).
From  radial velocity survey   CH stars are known  to be binaries. 
According to the models of  McClure (1983, 1984)  and   McClure \& Woodsworth 
(1990) the CH binaries have orbital characteristics consistent with the
presence of a white dwarf companion. Early type CH stars  are believed to 
have conserved the products 
of the carbon rich primary received through mass transfer and survived until
the present in the Galactic halo.
CH stars are not a homogeneous group of stars. They consist of two populations, 
the most metal-poor ones have a spherical distribution and the ones slightly 
richer in metals are characterised by a flattened ellipsoidal distribution 
(Zinn 1985).  The ratio of the local density of CH stars was found to be 
 ${\sim}$ 30\% of metal-poor giants  (Hartwick \& Cowley 1985).

 Many C-R stars also show a quite strong  G band of CH  in their 
spectra. Hence, based only  on the strength of the G-band of  CH it is 
not possible  to make a distinction between CH and C-R stars.
 In such cases the secondary 
P-branch head near 4342 \AA\, serves  as a more  useful indicator.
This is a well-defined feature in CH stars spectra in contrast
 to its appearance in C-R stars spectra (refer Fig 2, 3 of Goswami 2005).
Another  important  diagnostic  feature is  Ca I at 4226 \AA\, which  in 
case of CH stars is weakened by the overlying  CH band 
systems. In C-R star's spectra  this feature is  quite strong; usually  
the   line depth is  deeper 
than the depth of the  CN molecular band around 4215 \AA\,.  
These spectral characteristics allow for an identification of the CH  and 
C-R stars even at  low resolution.  

Abundances of neutron-capture  elements  can also be used as an useful 
indicator of spectral type. The abundances of  s-process elements  are nearly 
solar in C-R stars (Dominy 1984); whereas the CH stars  show significantly 
enhanced abundances of the s-process  elements relative to iron 
(Lambert et al. 1986, Green and Margon 1994). Although, the  C-R  as well as 
CH stars have warmer temperatures than those of C-N stars and blue/violet  
light is accessible 
to observation and atmospheric analysis, at low dispersion the 
narrow lines are difficult to estimate and elemental abundances can not  be
determined.   At low dispersion, therefore,  the `abundance criteria'
can not be used to  distinguish  the C-R stars from the  CH stars.
 Although the CH and 
C-R stars have similar range of temperatures, the distribution of CH stars 
place most of them in the Galactic halo. The large radial velocities,
typically $\sim$ 200 km s$^{-1}$  of the CH stars are indicative of their 
being halo objects (McClure 1983, 1984).
 
C-J stars spectra  are characterized by strong  Merrill-Sanford (M-S) bands 
ascribed to SiC$_{2}$  that appear in the wavelength region 4900 - 4977 \AA\,.
 The SiC$_{2}$ being a triatomic molecule, M-S bands are  expected to be the  
strongest in the coolest stars. SiC$_{2}$  and  C$_{3}$  have similar 
molecular structures and  in many C stars  C$_{3}$ molecule is believed to be
the cause of ultraviolet depression (Lambert et al. 1986).
 These bands are absent in the spectra of known  CH stars. A few warmer C-N stars 
are known to exhibit the presence of M-S bands in their spectra. 
 Strength of M-S bands are known to show a distinct correlation with  
carbon isotopic ratios; i.e., stars with higher $^{12}$C/$^{13}$C ratios show
weaker  M-S bands.  
 WZ Cas, V Aql \& U Cam  are a few exceptions that  have low $^{13}$C  
and  strong M-S bands (Barnbaum et al. 1996).
Strong C-molecular bands but a weak CH band characterize the class of
hydrogen deficient carbon stars. 

We have classified the program stars guided by  the above  spectral 
characteristics.  In the present  sample of ninety two stars the  spectra
of twenty two  objects are found  to not exhibit molecular bands of C$_{2}$.
 These objects are listed  in Table 2.
 Among the seventy    stars  that exhibit  strong carbon molecular bands
 (Table 1) thirty six  of them  are found to  show spectral characteristics 
of CH stars. 
These potential CH star candidates are listed in Table 3. 
  In the following  we  discuss the spectral characteristics
of the  individual  objects. 

\section{ Results and Discussions}

 The spectra of the objects are primarily  examined in terms of the following  
spectral characteristics.\\
1. The strength (band depth) of the CH band around  4300 \AA\,.\\
2. Prominance of the secondary P-branch head near 4342 \AA\,.\\
3. Strength/weakness  of the  Ca I feature at 4226 \AA\,.\\
4. Isotopic band depths  of C$_{2}$ and CN,  in particular the Swan bands
 of $^{12}$C$^{13}$C and $^{13}$C$^{13}$C near 4700 \AA\,.\\
5. Strengths of the other C$_{2}$ bands in the 6000 -6200 \AA\, region. \\
6. The $^{13}$CN band near 6360 \AA\, and the other CN bands across the
 wavelength range.\\
7. Presence/absence of the  Merrill-Sandford bands around 4900 - 4977 \AA\, 
region.\\
8. Strength of the  Ba II features at 4554 \AA\, and 6496 \AA\,.\\

The   membership of a star in a particular  group is established  from a 
differential analysis of the program stars spectra with the spectra
of  the comparison stars. Spectra of carbon stars  available in the 
low resolution spectral atlas of carbon stars of  Barnbaum et al. (1996)
are also consulted. 
In Figure 1 we reproduce
 the spectra of the comparison stars in the wavelength region
4000 - 6800 \AA\,; in figure 2 we show one example of HE stars spectra 
from the  present sample corresponding
to each comparison star's spectrum in figure 1.

\begin{figure*}
\epsfxsize=14truecm
\epsffile{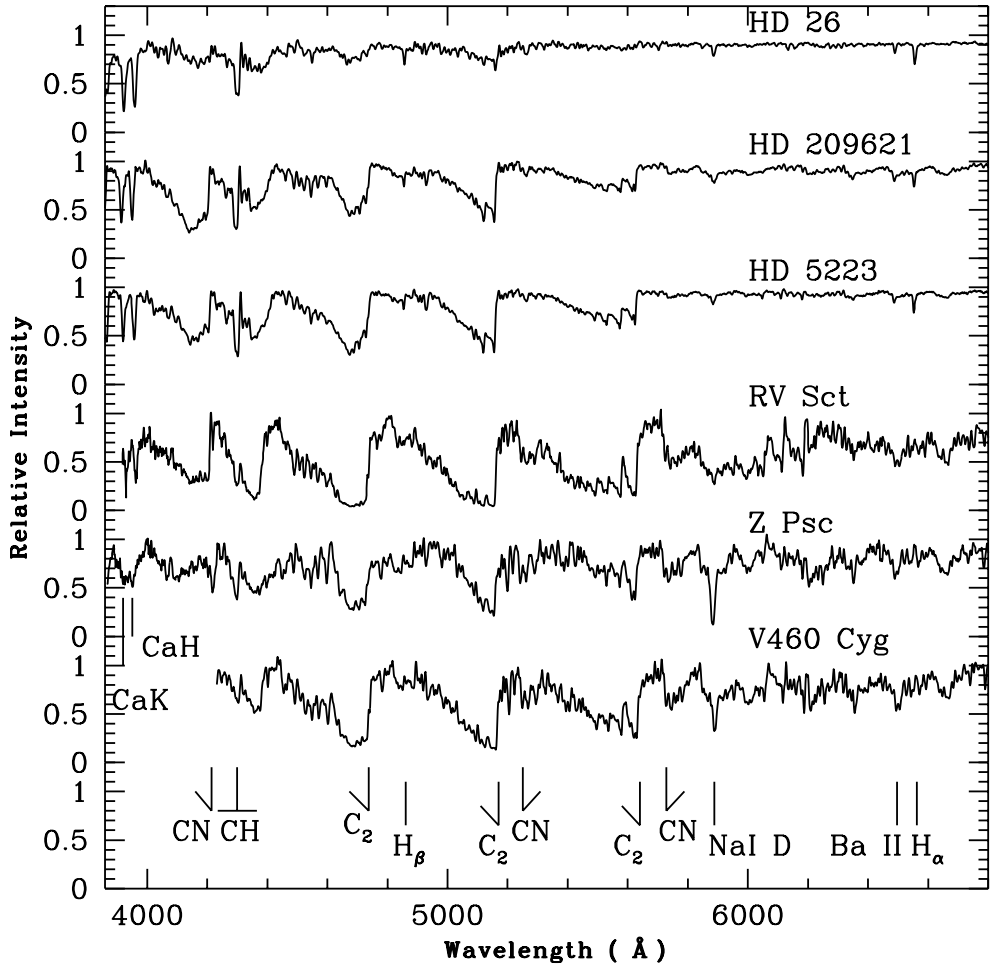}
\caption{ The spectra of the comparison stars in the wavelength 
region 3860 - 6800 \AA\,. Prominent features seen on the spectra are indicated.}
\label{Figure 1}
\end{figure*}

\begin{figure*}
\epsfxsize=14truecm
\epsffile{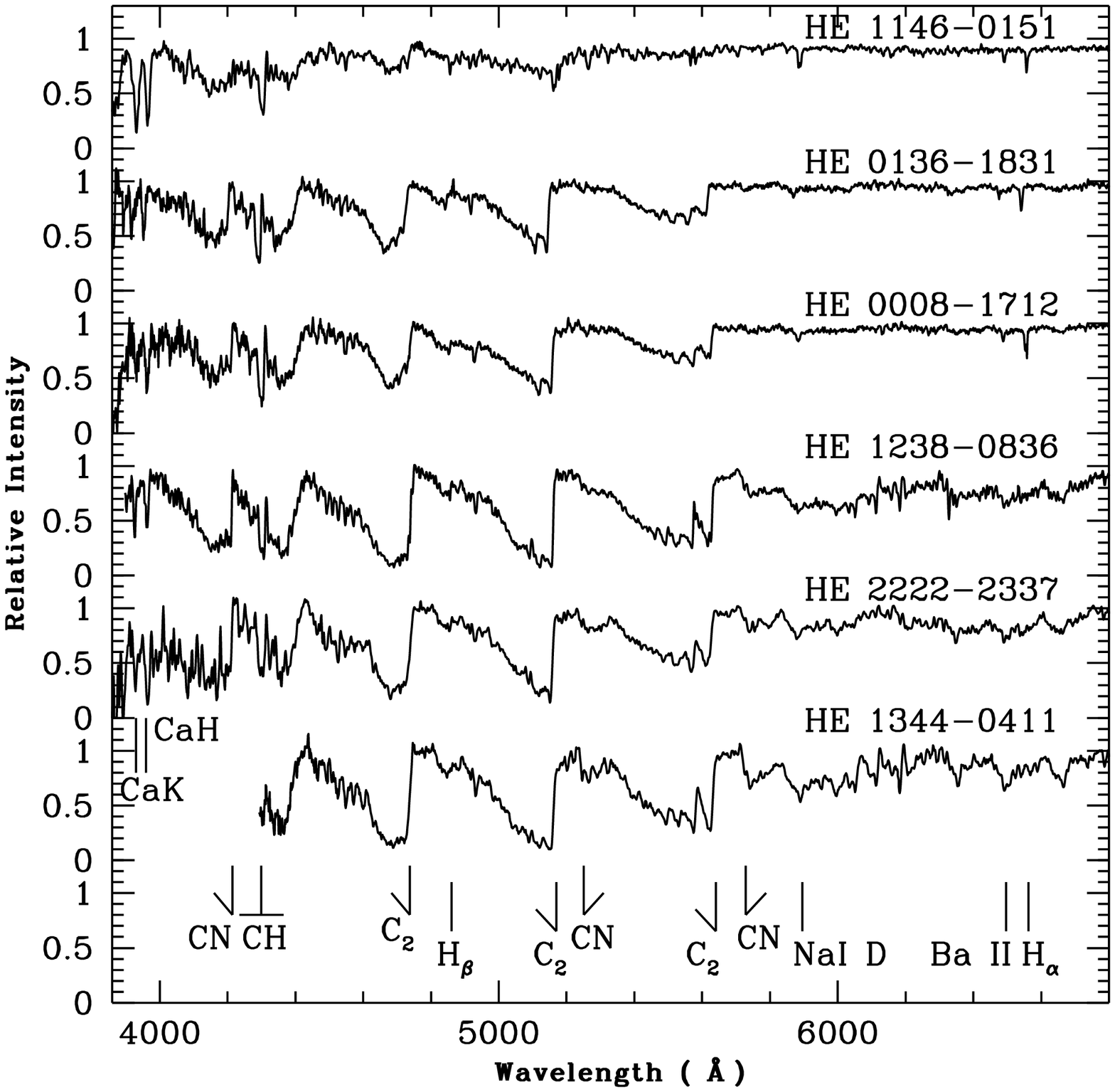}
\caption{ An example of each of the HE stars corresponding to the 
comparison stars presented in Figure 1, in the top to bottom sequence,
in the wavelength region 3860 - 6800 \AA\,. The locations  of the
prominent features  seen in the spectra are marked on the figure.
}
\label{Figure 2}
\end{figure*}

\subsection { Location of the  candidate CH stars on (J-H) vs (H-K) plot}

We have  used JHK photometry as  supplementary diagnostics for stellar 
classification. Figure 3 shows the locations of the candidate CH stars 
listed in Table 3  on the (J-H) vs (H-K) plot.
 2MASS JHK measurements of the stars  are available on-line at
http://www.ipac.caltech.edu/.
The thick box on the lower left represents the location of CH stars and the
thin box on the upper right represents the location of C-N stars (Totten et
al. 2000). 
 Except two lying outside (shown with open squares in figure 3), the  locations 
of the   candidate CH  stars (shown with open circles) are well within the CH 
box. This  supports  their identification  with the class of  CH stars. 
Location of the  comparison  CH stars
HD~26, HD~5223 and  HD~209621, C-R star  RV Sct, C-N stars V460~Cyg 
and  Z PSc,  are  shown by solid squares on the (J-H) vs (H-K) plot.
As described in the next section, we have  also used  2MASS JHK photometry 
 to determine the effective temperatures of the objects.

\begin{figure*}
\epsfxsize=14truecm
\epsffile{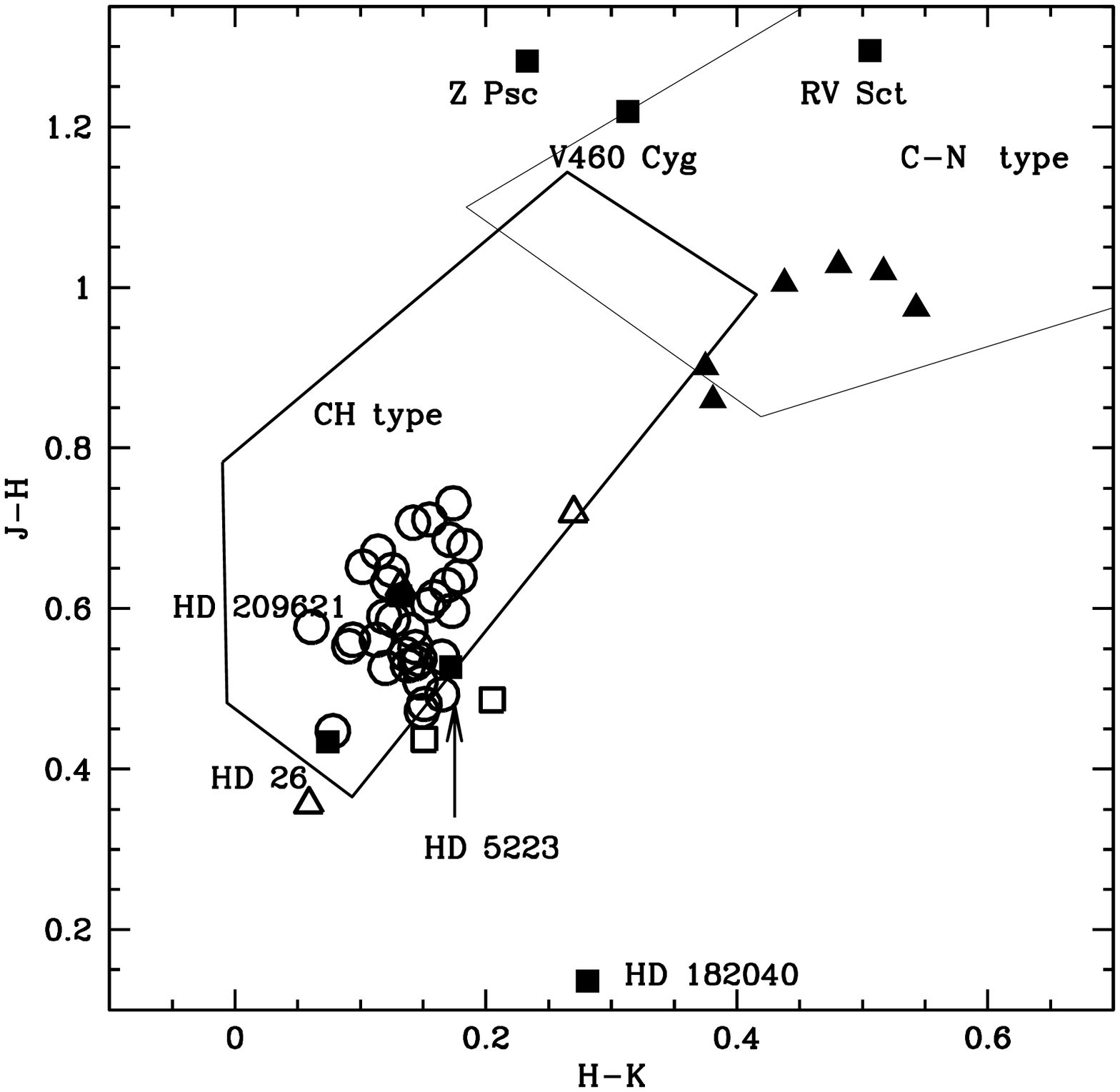}
\caption{
 A two colour  J-H versus H-K diagram of the candidate CH stars.
The thick box on the lower left represents the location of CH stars and the
thin box on the upper right represents the location of C-N stars (Totten et
al. 2000).  Majority of the  candidate CH stars
listed in Table 3  (represented by open circles) fall well
 within the CH box. The positions of the two outliers are shown with open
squares. C-N stars  found in our sample are represented by solid 
triangles.  The location of the 
comparison stars are labeled and marked with solid squares. Location of the 
three dwarf carbon stars are indicated by open triangles.}
\label{Figure 3}
\end{figure*}

\subsection{ Effective temperatures of the program stars}
Semiempirical temperature calibrations offered by Alonso et al. 
(1994, 1996, 1998)
are used to derive priliminary temperature estimates of the program stars.
These authors used the infrared flux method to measure temperatures for a 
large number of lower main sequence stars and subgiants  to derive
 the calibrations. The calibrations relate T$_{eff}$ with Stromgren 
indices as well as [Fe/H] and colours (V-B), (V-K), (J-H) and (J-K). The
calibrations  hold within a temperature and
metallicity range 4000 ${\le}$ T$_{eff}$ ${\le}$ 7000 K and
-2.5 ${\le}$ [Fe/H] ${\le}$ 0 . The estimated uncertainty in T$_{eff}$ 
arising from different sources is ${\sim}$90 K (Alonso et al. 1996).
Alonso et al.  derived the T$_{eff}$ scales using   photometric data 
measured on  TCS system; 2MASS  JHK photometric data  are therefore
converted to the  TCS system using the conversion relations 
of  Ramirez and Melendez (2004).
Estimation of T$_{eff}$ from  (J-H) \&
 (V-K) temperature relations involve a metallicity term; (J-K) calibration
relation is independent of metallicity.
We have estimated the effective temperatures using  adopted metallicities
shown in parenthesis in Table 4.
 (B-V) calibration,  normally used in case of normal stars  is  not 
considered as in the case of carbon stars the colour 
 B-V depends on the 
chemical composition and metallicity in addition to T$_{eff}$. B-V colour 
often gives a  much lower value than the actual surface temperature of the 
star due to the effect of CH molecular  absorption in the B band.
We have  assumed that the effects of reddening on the measured colours
are  negligible.

\subsection{ Isotopic ratio $^{12}$C/$^{13}$C from molecular band depths}

  Carbon isotopic ratios  $^{12}$C/$^{13}$C,  widely used as  mixing 
diagnostics  provide an important  probe of stellar evolution.  
These ratios  measured on low resoltuion spectra do  not give accurate 
results  but provide  a fair indication of evolutionary states of the objects.

We have estimated  these ratios, whenever possible, using  the molecular band 
depths  of (1,0) $^{12}$C$^{12}$C ${\lambda}$4737 and
(1,0) $^{12}$C$^{13}$C ${\lambda}$4744.
For a majority of the  candidate CH  stars,
the ratios $^{12}$C/$^{13}$C are found to be  ${\le}$ 10.
These ratios  for  HD 26, HD 5223 and HD 209621 are respectively
5.9, 6.1  and 8.8 (Goswami 2005).

Our estimated ratios of  $^{12}$C/$^{13}$C indicate that most of the candidate
CH stars belong to the `early-type' category.
The low carbon isotope ratios imply that, in a binary system,
the material transferred from the now unseen companion has been mixed into
the CN burning region of the CH stars or constitute  a minor fraction of the
envelope mass of the CH stars. Low  isotopic ratios are  typical of stars
on their first ascent of the giant branch.
The $^{12}$C/$^{13}$C ratios and the total carbon abundances decrease 
due to the convection which dredges up the products of internal CNO cycle to 
stellar atmosphere as ascending RGB. If it reaches AGB stage,  fresh $^{12}$C 
may be supplied from the internal He burning layer to stellar surface leading 
to an increase of  $^{12}$C/$^{13}$C ratios  again. 

\subsection{ Spectral characteristics of the  candidate CH  stars } 

\begin{figure*}
\epsfxsize=14truecm
\epsffile{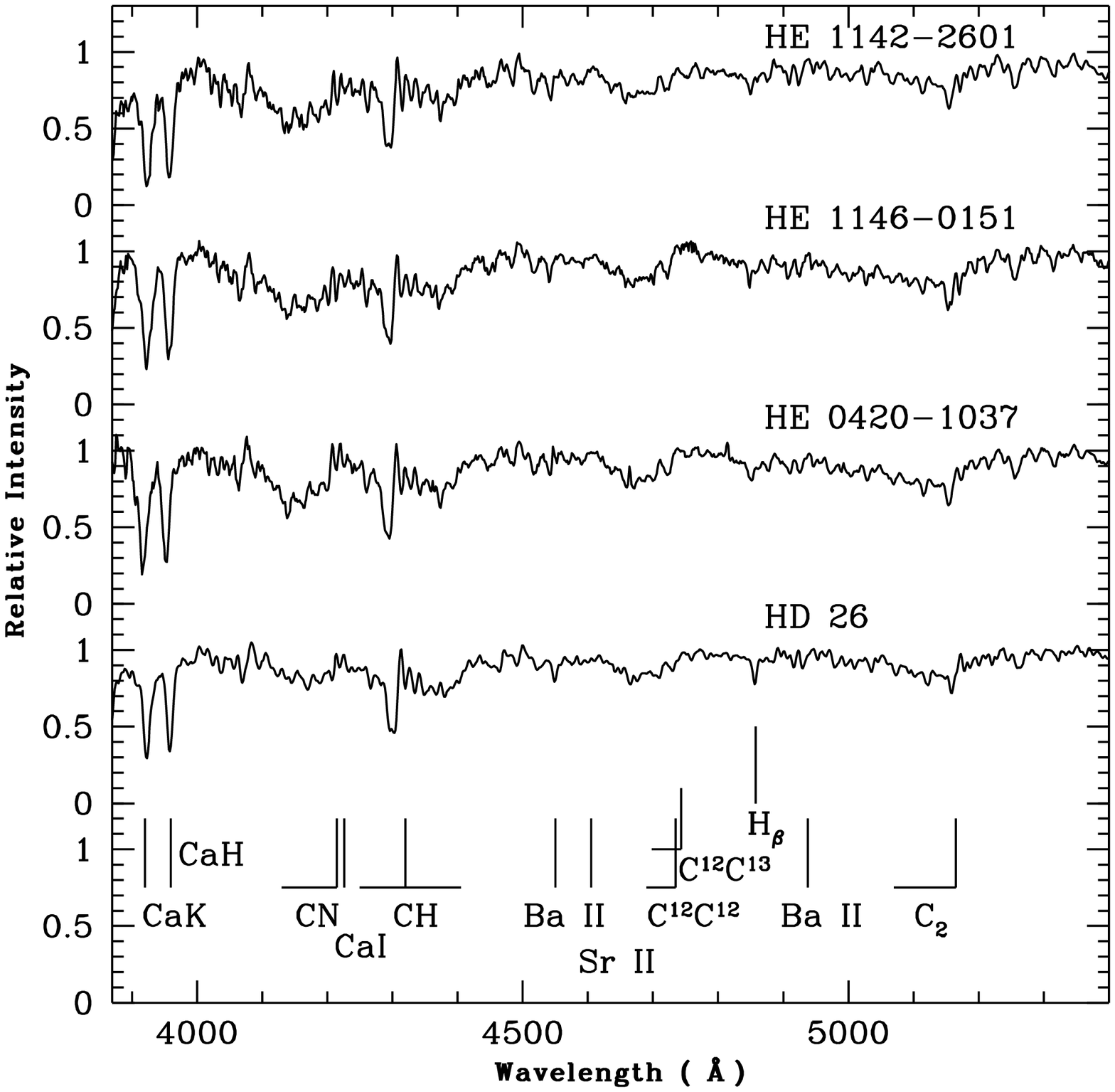}
\caption{ A comparison of the spectra of three HE stars in the
wavelength region 3870 - 5400 \AA\, with the spectrum of the comparison star
HD~26. Prominent features  noticed in the spectra  are marked on the figure.
}
\label{Figure 4}
\end{figure*}

{\bf HE~0420-1037, ~1102-2142, ~1142-2601, 
~1146-0151, ~1210-2636, ~1253-1859, ~1447+0102.~~ }
The spectra of these objects resemble closely the spectrum of HD~26.
HD~26 is a known  classical CH star with effective temperature 
5170 K and log $g$ = 2.2 (Vantures 1992b). 
 The temperatures of these objects as measured using JHK
photometric data range from 4200 to 5000 K.
 The locations of these objects are well within the CH box in figure 3.
In figure 4, we show as an example, a comparison of the spectra of three 
objects HE~0420-1037, HE~1142-2601  and HE~1146-0151 with the spectrum 
of HD~26.
With marginal differences in the strengths of the molecular features,  the 
spectra of these 
three objects show more or less a good match with their counterparts in HD~26. 
The  CN  band around  4215 \AA\,  and the  C$_{2}$ band around  5165 \AA\, in 
the spectra
of HE~0420-1037 and HE~1146-0151 are marginally  stronger; the Na I D  
feature also 
appears stronger. The features due to Ca K and H appear with similar strengths.
In the spectrum of HE~1142-2601, the CN  band depth around  4215 \AA\,
  and Ca II K and H line depths are deeper than their counterparts in  HD~26.
The Ca I line  at 4226 \AA\,  is detected with line depth weaker than the band 
depth around 4215 \AA\,.  In HD~26,  the  Ca I 4226 \AA\, feature  is 
not detected.
The object   HE~1253-1859  also have very similar spectrum with that of HD~26.
The  CN band around 4215 \AA\, and  carbon 
isotopic band around 4730 \AA\, are   stronger,  but the CH band, Ca II K and H 
features are of similar strengths. The molecular bands around 5165 \AA\,  
and 5635 \AA\,
 show an exact  match.  The lines due to Na  I D, Ba II at 6496 \AA\, and
 H$_{\alpha}$ are seen equally strongly as in HD~26. 
The Ca I feature at 4226 \AA\, could not be detected and the
 secondary P-branch head around 4342 \AA\, seems to be marginally 
weaker.  
  In the  spectrum  of   HE~1102-2142 
the molecular  C$_{2}$ bands  around 4735, 5165 and 5635 \AA\,  are  slightly 
deeper than those in HD~26. The  CN band around 4215\AA\, and the CH band 
around  4310 \AA\,  
also appear marginally stronger in the spectrum of  HE~1102-2142. The
H$_{\alpha}$  feature and the Ba~II line at 6496 \AA\, are  marginally weaker;
the feature due to Na I D
appears with similar strength as in  HD~26.  H$_{\beta}$ at 4856\AA\,
is  clearly seen. The Ca I line at 4226 \AA\, is weakly detected.
In the spectrum of HE~1210-2636, the CN band around 4215 \AA\, and 
the secondary P-branch head around 4342 \AA\, appear slightly stronger than 
in HD~26. The C$_{2}$ molecular band around 5165 \AA\, is also  slightly 
weaker.
The  carbon isotopic band  around 4733 \AA\, is marginally detectable. 
Ca I  line at 4226 \AA\, could  not be 
detected. Features due to Ca II K and  H and H$_{\alpha}$ are  clearly detected.
 In the spectrum of  HE~1447+0102, the CN band around  4215 \AA\, is almost
absent. Strong well defined features due to Ca II K and
H are seen.  Molecular bands around 4733,  5165, and  5635 \AA\,
 are distinctly seen to be stronger than their counterparts in HD~26. 
In the redward of 
5700 \AA\,  no  molecular  bands are  detected. We assign these objects to 
the CH group. \\

\begin{figure*}
\epsfxsize=14truecm
\epsffile{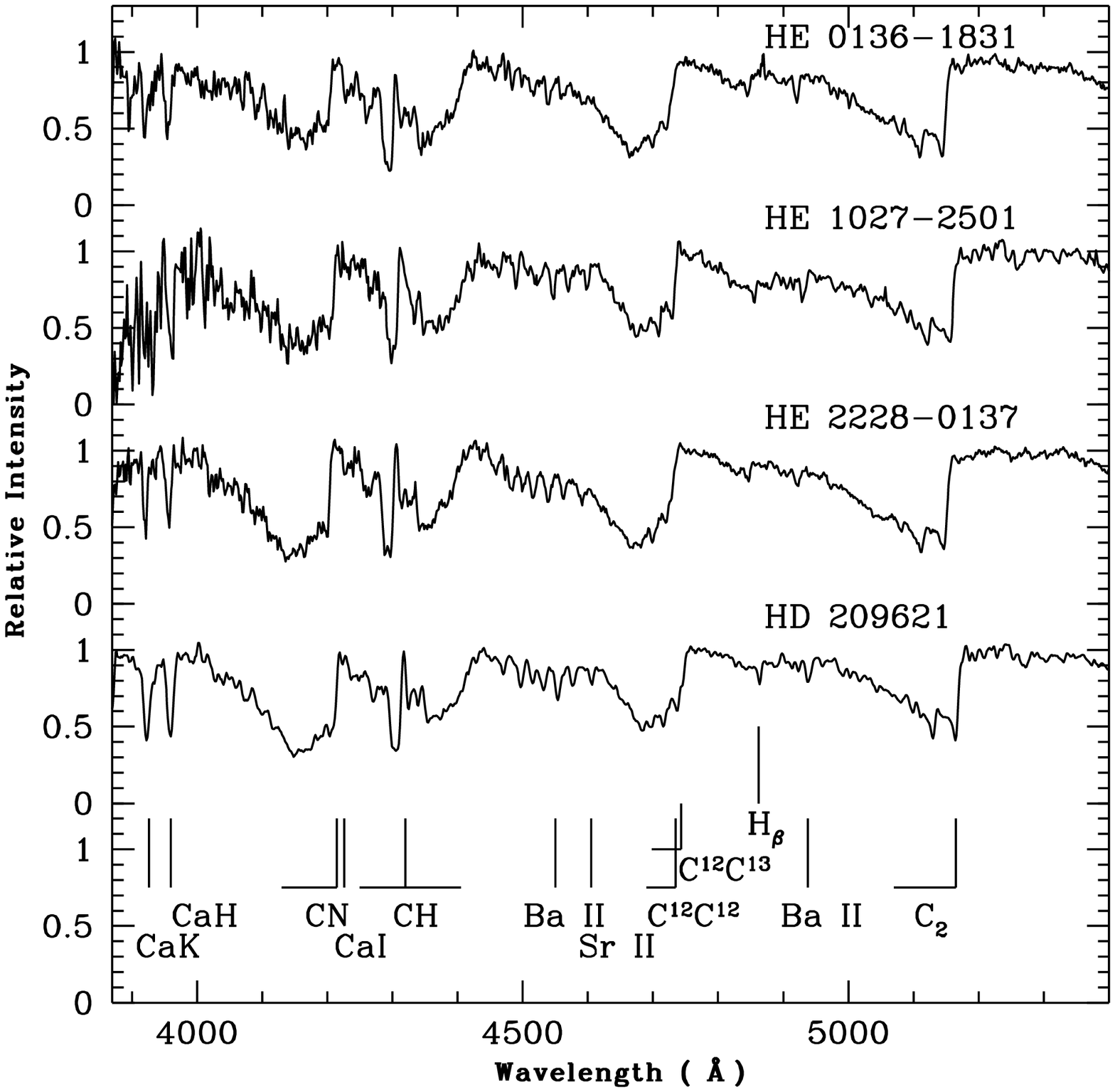}
\caption{ A comparison of the spectra of three HE stars in the
wavelength region 3870 - 5400 \AA\, with the spectrum of the comparison star
HD~209621. Prominent features  noticed in the spectra  are marked on the figure.
}
\label{Figure 5}
\end{figure*}
{\bf HE~0136-1831, ~0308-1612, ~1027-2501, ~1051-0112, ~1119-1933, ~1120-2122, 
~1123-2031, ~1145-1319,  ~1205-2539,  ~1331-2558, ~1404-0846, 
~1425-2052, ~1525-0516, ~2114-0603, ~2211-0605, 
~2228-0137, ~2246-1312.~~ } 
The spectra of these objects resemble closely  the spectrum of HD~209621,
a  classical CH star  with effective temperature  ${\sim}$  4400 K 
(Tsuji et al. 1991). The  effective temperatures estimated for this set 
of objects  using JHK 
photometry range from 3948 K (HE~2114-0603) to 4675 K (HE~1119-1933). Their
locations on the J-H vs H-K plot are well within the CH box in figure 3.
Three examples,  HE~0136-1831, HE~1027-2501 and HE~2228-0137  from this 
set are  shown in figure 5 together  with the spectrum of HD~209621.
In the spectrum of HE~0136-1831, the CN band around 4215 \AA\, is 
marginally weaker and the carbon molecular 
bands around 4733 and 5635 \AA\,  are marginally stronger than those 
in HD~209621. All other features show a good match.  The spectrum of 
HE~1027-2501  also shows a  close match with the spectrum of HD~209621. 
Except for the molecular bands around 4733, 5165 and 5635 \AA\, that appear 
marginally weaker in the spectrum of HE~2228-0137 the spectrum of this
 object  bears a close resemblance with the spectrum of  HD~209621.
The spectrum of  HE~0308-1612  shows  weaker molecular bands around 5165 \AA\, 
 and 5635 \AA\,. The CN band around 4215 \AA\, and  the carbon 
isotopic band around 4730 \AA\, are  of similar strengths. The CH band around 
4300 \AA\, and  Ca II K  and H  features are of similar depths.  The  lines 
due to Na I D, Ba II at 6496 \AA\, and  H$_{\alpha}$ are clearly noticed.
The Ca I feature at 4226 \AA\, could not be detected. 
The spectrum of  HE~1051-0112 shows a weaker CN  band around 4215 \AA\,. The 
G-band 
of CH appears with  almost the  same strength   as in HD~209621. The secondary 
P-branch head around 4342 \AA\,  and  the bands around  4730 and 5635 \AA\, 
are  relatively stronger. Features due to Ca II K and H are barely detectable 
in the spectrum of this object. The molecular  band around 5165 \AA\, shows an 
exact match with its counterpart in HD~209621. The features due 
to H$_{\alpha}$ and Na I D  are detected distinctly; the  Ba II feature at
  6496 \AA\, is  marginally detected.

In the spectrum of HE~1119-1933, Ca II K and  H
appear marginally stronger. The molecular band around 5365\AA\, appears 
marginally 
weaker than in HD~209621.  The spectra of HE~1120-2122 and HE~1123-2031 are 
very similar, both  exhibit a weaker CN band around 4215 \AA\,.
 All other features show a  good match with their counterparts in the spectrum 
of HD~209621.
Ca II K and H appear with almost  the same strength in the spectrum of 
HE~1120-2122 as in HD~209621. Ca I line at 4226 \AA\, is not detectable.
The spectra of HE~1123-2031, HE~1145-1319, HE~1205-2539, HE~1331-2558 
 are  noisy shortward of 4100 \AA\,
and  the lines due to Ca II K and H  could not be clearly detected.
Ca I line at 4226 \AA\, is not detectable in these spectra.
Features due to Na  I D, H$_{\alpha}$, and Ba II at 6496 \AA\, are detected. 
 In the spectrum of HE~1145-1319, C$_{2}$ molecular bands around 5635 and 
4733 \AA\,  are marginally deeper than those in HD~209621.  
The features redward of 4200 \AA\, in the spectrum of HE~1205-2539 show a good 
match with those in HD~209621. The molecular features in the spectrum 
 of HE~1331-2558 are  also of similar strenths with those in HD~209621. 
Except for the G-band of CH, all other molecular features are weaker in the 
spectrum of HE~1404-0846. The features due to Ca II K  and H are marginally 
stronger. The spectrum of HE~1404-0846 is noisy at the blue end.

In the spectrum of  HE~1425-2052 the G-band of CH  and  the C$_{2}$ band 
around 5635  \AA\,  are mildly stronger.
The rest of the features show a close  match   with their counterparts 
in HD~209621.
The  Ba II feature at 6496 \AA\,  appears with equal intensity  as that of
H$_{\alpha}$ feature. The feature due to Na  I D is clearly detected.
In the spectrum  of HE~1525-0516, H$_{\alpha}$ and  Na I D features
 are detected with almost equal strength as  in HD~209621. 
The spectrum of HE~2114-0603 shows a remarkably close  match with
the features in HD~209621 including those longward of 5700 \AA\,. However the
features due to Ca II K  and H that are seen very distinctly in the spectrum 
of HD~209621 could not be detected in the spectrum of HE~2114-0603;
the spectrum  is noisy blueward of 4000 \AA\,.
In the spectrum of  HE~2211-0605  the molecular
 features are weaker than their counterparts in  HD~209621 but stronger 
than those in HD~26. The CH band  matches  exactly with the one in HD~209621.
 Ca II K and H appear with  almost  equal strengths as in HD~209621.
The spectrum of HE~2246-1312 shows a  weaker molecular band  around
CN 4215 \AA\,. Other molecular bands appear with  almost of equal strengths 
as their counterparts in HD~209621.  The spectrum 
blueward of 4100  \AA\, is  noisy and 
 Ca II K and H features could not be detected as  well defined features.
The spectrum obtained in  september 2008 has a better signal. \\

\begin{figure*}
\epsfxsize=14truecm
\epsffile{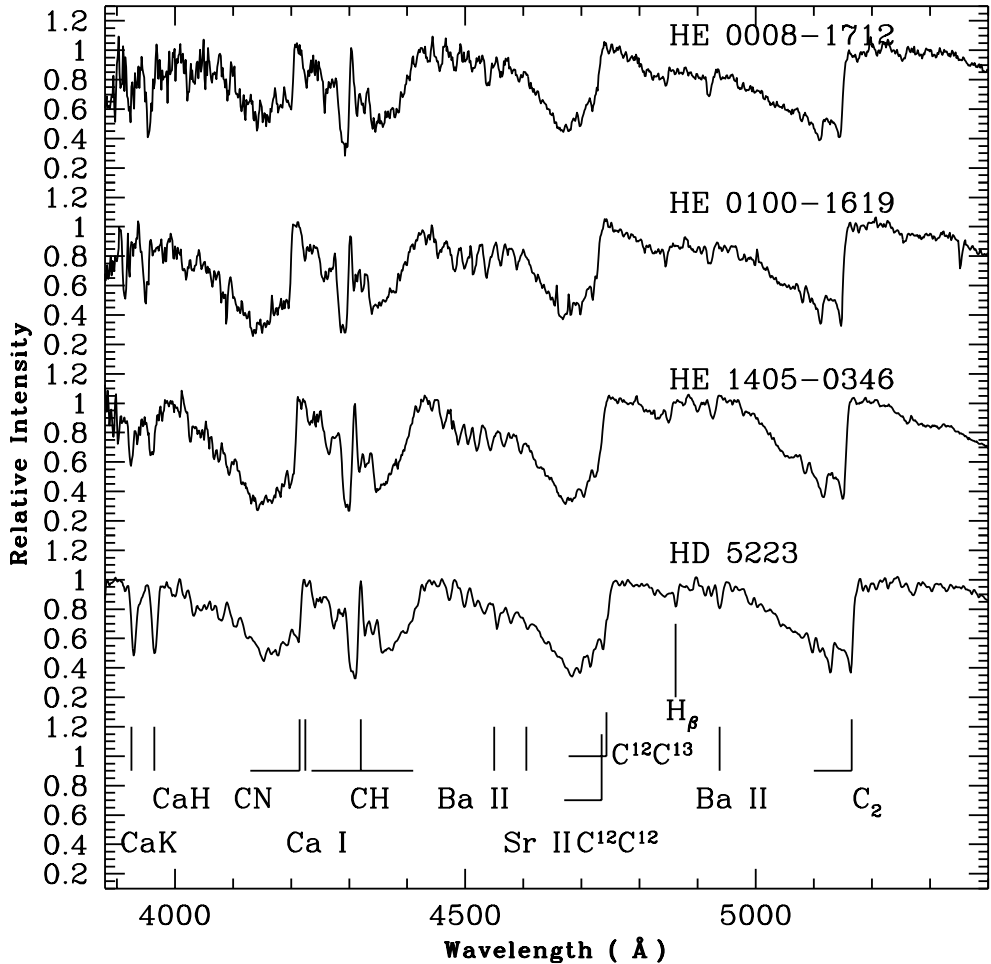}
\caption{ A comparison of the spectra of three HE stars in the
wavelength region 3880 - 5400 \AA\, with the spectrum of the comparison star
HD~5223. Prominent features  noticed in the spectra  are marked on the figure.
}
\label{Figure 6}
\end{figure*}

 {\bf HE~0008-1712, ~0052-0543, ~0100-1619, ~0225-0546, ~1045-1434, 
~1110-0153, ~1157-1434, ~1228-0417, ~1405-0346, ~1410-0125, ~1431-0755,
 ~1440-1511.~~ }
The spectra of these objects closely resemble the spectrum of HD~5223, 
a well-known classical CH star with effective temperature ${\sim}$ 4500 K,
 log $g$ = 1.0 and metallicity [Fe/H] = $-$2.06 (Goswami et al. 2006).
The effective temperatures of these objects  derived from J-K colour range 
from  about 3924 K (HE~0052-0543)  to 4795 K (HE~1228-0417).
Except for the two outliers  HE~1045-1434 and  HE~1228-0417 (represented 
with open squares)  the locations  of this set of 
objects are well within the CH box in figure 3.

A comparison of  the spectra of HE~0008-1712, HE~0100-1619 and HE~1405-0346 
with the spectrum of HD~5223 is shown in figure 6.
The Ca I line at  4226 \AA\, is not detectable in any of these spectra.
The  CH band as well as other molecular bands  show a very good match.
The features due to Ca  II K and H  are seen with equal strength as in HD~5223.
The CN band around 4215 \AA\, in HE~0100-1619
is slightly deeper. The bands longward of 5635 \AA\, are also  marginally 
deeper.
This object HE~0100-1619 is also mentioned as a CH star in 
Totten et al. (2000). Heliocentric radial velocity of HE~0100-1619 as reported
  by Bothun et al. (1991) is $-$142 km s$^{-1}$.

 The spectra of  HE~0052-0543 and HE~1110-0153 
show  stronger molecular bands than their counterparts in HD~5223.
In the spectrum of HE~0225-0546 the molecular  bands are  marginally stronger 
than  in HD~5223. The molecular  features  above 5700 \AA\, seen  in these two
spectra  are barely noticed in the spectrum of HD~5223.
The spectrum of HE~1157-1434 also show a good match with the spectrum of
HD~5223 except for the molecular band  around 5635 \AA\, which is distinctly
 weaker in its spectrum.
The molecular features redward of  5700 \AA\, are  also noticed
weakly  in the spectrum of this object.
 The spectrum of HE~1405-0346 shows a stronger  CN  band around
4215 \AA\, as well as a stronger carbon molecular  band around 5635 \AA\,. 
The secondary P-branch head near 4342 \AA\, is also stronger
than its counterpart in HD~5223. Other molecular bands around 4733 and  
5165 \AA\,   show a  good match.  Ca  II K and H are seen as strongly as 
in HD~5223.
 The effective temperature of the object from J-K colour is 4391 K, slightly 
lower than the  effective temperature of HD~5223. 
In the spectrum of HE~1410-0125 the molecular features are slightly shallower
than their counterparts in HD~5223. The CH band depth is however of  similar 
strength.  The feature at  Ca I 4226 \AA\, is absent; the features due to 
 Ca II K and H  are of similar strengths. The CN band around 4215 \AA\,  
matches  well with the CN feature in HD~5223.  The  radial velocity
of this object  as quoted by Frebel et al. (2006) is  +80 Km s$^{-1}$. 
The effective temperature  estimated using
(J-K) calibation returns a value 4378 K for this object.

The spectrum of HE~1431-0755 is noisy blueward of 4000 \AA\,; the features of
  Ca II K  and H   could not be   detected. The CH band around 4310 \AA\,  and
 the  CN band around  4215 \AA\, appear
 slightly stronger than their counterparts in HD~5223. 
Other C$_{2}$ molecular  bands present in the spectrum are narrower than their
counterparts in HD~5223. The spectrum redward of 5700 \AA\, shows molecular 
features that are barely noticed in the spectrum of HD~5223.
The spectrum of HE~1440-1511 shows molecular bands with almost equal depths
with  those in HD~5223. Ca II  K and H features are however stronger than 
their counterparts in
HD~5223.  The spectrum shows a  distinctly stronger feature due to Na I D. 
Features of 
H$_{\alpha}$ and Ba II at 6496  \AA\, are of equal strengths. The spectrum  
redward  of 5700 \AA\, shows a good match.

In the spectrum of HE~1228-0417 the  Ba II feature at 6496  \AA\, 
and H$_{\alpha}$ are
seen with equal strengths as in HD~5223. The part of the spectrum redward of
5700 \AA\, shows a very good match. The Ca I line at 4226 \AA\, is not detected.
The feature due to Na I D  is clearly detected. Other carbon molecular features
around 4730, 5165, and 5635 \AA\, appear marginally weaker than their 
counterparts in HD~5223. The G-band of CH appears with almost equal 
strength but  the CN band around 4215 \AA\, is marginally weaker than its 
counterpart  in HD~5223.
 The effective temperature of the object estimated using  J-K colour 
calibration is 4795 K, higher than the effective temperature of 
HD~5223 ${\sim}$ 4500 K (Goswami et al. 2006). The location of this  object
outside the CH box is not obvious from its low resolution spectra.\\

\begin{figure*}
\epsfxsize=14truecm
\epsffile{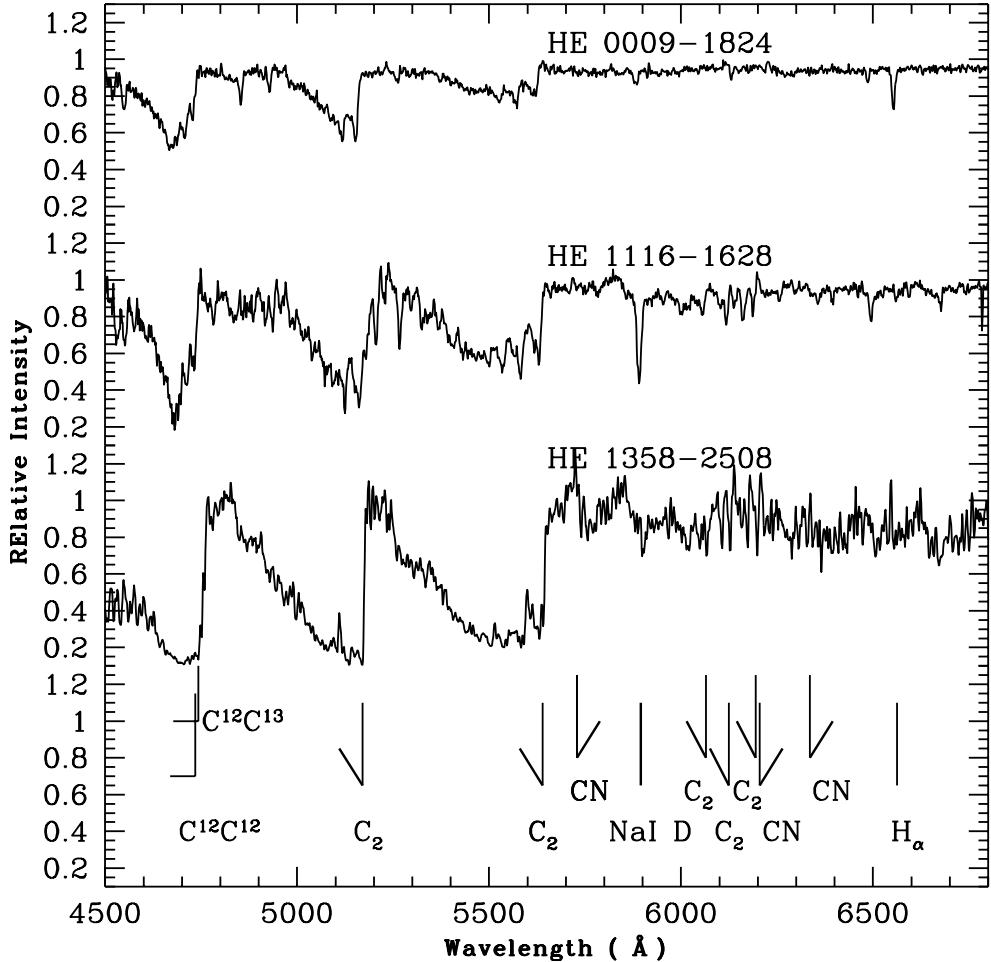}
\caption{ The spectra of three  dwarf carbon stars in the
wavelength region 4500 - 6800 \AA\,. Prominent features  noticed in the 
spectra  are marked on the figure.
}
\label{Figure 7}
\end{figure*}

{\bf HE~0009-1824, ~1116-1628, ~1358-2508.~~} 
 The spectra of these objects are illustrated in figure 7. 
These three objects are known  dwarf carbon stars.
The effective  temperatures of HE~0009-1824, HE~1116-1628, HE~1358-2508
  as estimated from (J-K) calibration are respectively 5530 K, 4224 K and 
3623 K. As expected, the molecular band depths are the  strongest  
in HE~1358-2508, 
the  coolest of the three objects; and weakest in HE~0009-1824.
In the  spectrum of HE~0009-1824 the CN band around 
4215 \AA\, is completely missing.
The features due to  Ca II K  and H   as well as the CN band near 3880 \AA\, 
are detected. The G-band of  CH  is strong but not as strong 
as it appears  in CH stars. 
The secondary P-branch head near 4342 \AA\, is seen distinctly. Apart from 
the absence of the  CN band around  4215 \AA\,  the 
spectrum of this object looks somewhat  similar to the spectrum
 of HD~209621. 
The distance of this object as reported by Mauron et al. (2007) is 300 pc. 

The spectra of HE~1116-1628 and HE~1358-2508  show characteristics of C-R star 
RV Sct with marginal differences in the  molecular band depths. 
In the spectrum of HE~1358-2508, the CH  band is marginally stronger 
than in  RV Sct. The C$_{2}$ molecular  bands are  stronger in the spectrum 
of this object. The CN  band around 4215\AA\, is  clearly detected.
Ratnatunga (1983) first proposed  this object   HE~1116-1628 to be a 
dwarf carbon
star. This object  is  also present in the list of dwarf carbon stars
of  Lowrance et al. (2003). Mauron et al. (2007)  reported  the proper motions
 in ${\alpha}$ and
${\delta}$ and their respective 1${\sigma}$ errors in mas yr$^{-1}$ as
$-23.5 \pm 6.7$ and $+29.8 \pm 4.6$. 
  The distances of HE~1116-1628 and HE~1358-2508 
as reported by Mauron et al. (2007) are  170 pc and 270 pc respectively.
 Totten and Irwin (1998)  reported a radial velocity of $-69$ km s$^{-1}$
 for the object  HE~1116-1628.
All the  three objects have total proper motion ${\mu}$ 
${\ge}$ 30 mas yr$^{-1}$ (Mauron et al. 2007).

  The locations  of the three dwarf carbon stars are 
 indicated by open triangles in figure 3. Location of HE~0009-1824
is on the left below the CH box, the location of HE~1358-2508 is on the
right edge of the CH box and the location of HE~1116-1628 is  found to be 
well inside the CH box. 
 Dwarf carbon stars have anomalous infrared colours (Green et al. 1992 and 
Westerlund et al. 1995). In the conventional two colour JHK diagram
the locus of dwarf-carbon-stars colours is away  from the normal carbon-star
locus. The locus defined by  dwarf carbon stars is bounded by 
(J-H) ${\le}$ 0.75
and (H-K) ${\ge}$ 0.25 (Westerlund et al. 1995). This condition is 
satisfied by HE~1358-2508; however
HE~0009-1824, and  HE~1116-1628 both have (H-K) colours less than the lower 
limit of 0.25 mag set for dwarf carbon stars.\\

\begin{figure*}
\epsfxsize=14truecm
\epsffile{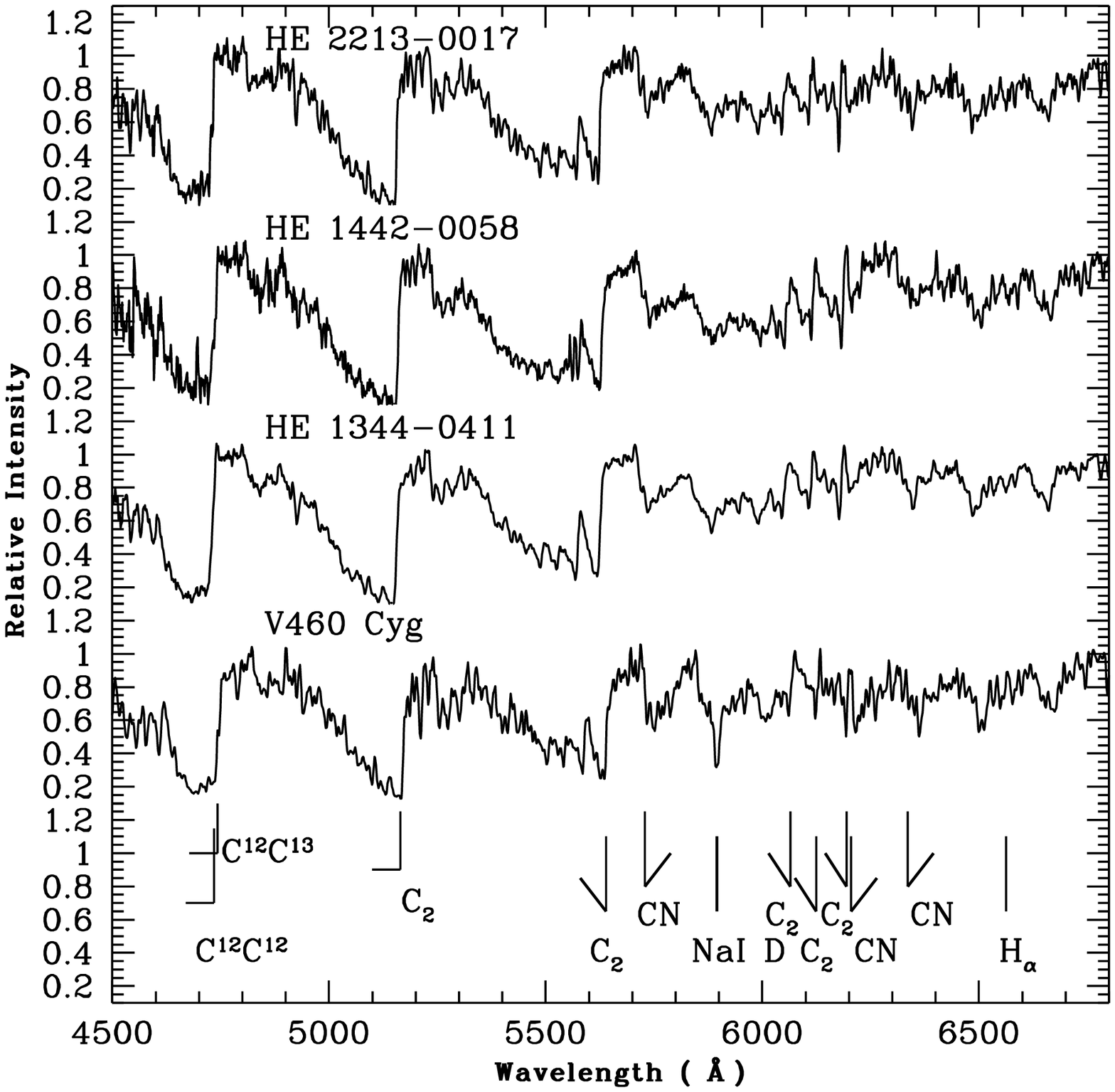}
\caption{ A comparison of the spectra of  the candidate C-N stars with the
 spectrum of V460 Cyg in the wavelength region 4500 - 6800 \AA\,.
 The bandheads of the prominant molecular bands, Na I D and H$_{\alpha}$ 
are marked on the figure.
}
\label{Figure 8}
\end{figure*}

{\bf Candidate C-N stars:   HE~0217+0056, ~0228-0256, ~1019-1136, 
~1344-0411, ~1429-1411, ~1442-0058, ~2213-0017, ~2225-1401.~~}  \\
The spectra of these objects show a close resemblance with the spectrum of the
C-N star Z PSc with similar strengths of CN and C$_{2}$ molecular
bands across the wavelength regions. 
In figure 3, the objects  HE~0217+0056, HE~1019-1136,  HE~1344-0411, 
HE~1429-1411, HE~1442-0058 and    HE~2213-0017 represented by solid triangles
 lie  well within the CN box.
The spectra  have  low flux   below  about 4400 \AA\,.
 The spectrum  of HE~1429-1411 is  similar to that  of Z PSc's spectrum,
 except that the  CN band around 4215 \AA\, is  marginally weaker in this star. 
The  CH band is weakly detected in the spectrum of HE~1116-1628.  The molecular
bands near  4735, 5135 and 5635  \AA\, are  noticed distinctly.   The feature
due to Na I D is strongly
detectable.  The Ba II line  at  6496 \AA\, is detectable
 but the  H$_{\alpha}$ feature could not be detected. 
HE~1127-0604 has low flux below about 4200 \AA\,. The  CH band and 
 C$_{2}$  molecular bands 
around 4735, 5165, 5635 \AA\, are detected. All the  features  in the spectrum 
are weaker than their counterparts in Z Psc. While features of 
Ca II K and H are  detected, the CN band around 4215 \AA\, is  not clearly seen.
The spectrum of HE~2225-1401 has low flux below about  4700 \AA\,.
The strong C$_{2}$ molecular  bands around 5165, 5635 \AA\, appear
 stronger than their counterparts  in Z Psc. The spectrum of  HE~2225-1401 
although have spectral characteristics of   C-N stars, its location in figure 3
 is  not within the CN  box. The  spectra of the objects
 HE~2213-0017, HE~1442-0058, HE~1344-0411  compare closest to the
spectrum of C-N star V460 Cyg as illustrated in figure 8. 

The objects  HE~0217+0056, HE~1019-1136, HE~1442-0058, HE~2213-0017 and 
HE~2225-1401 are  also  mentioned  as N-type stars   in the 
APM survey of cool carbon stars in the Galactic halo (Totten \& Irwin 1998).
Totten et al. (2000) have provided proper motion measurements for 
these objects.  The distances measured
by these authors  assuming an average M$_{R}$ = $-$3.5 for these objects  are 
respectively 24, 16, 43, 29 and 24 kpc. Heliocentric radial velocities  
estimated by
Totten \& Irwin (1998) for these objects are respectively 
$-142 \pm 3$, $126 \pm 4$, $37 \pm 4$, $-44 \pm 3$, and $-113 \pm 5$ 
 km s$^{-1}$.
Heliocentric radial velocity of   HE~0228-0256  is $-72$ km s$^{-1}$
 (Bothun et al. 1991).\\

{\bf HE~0945-0813, ~1011-0942,  ~1127-0604, ~1205-0521,  ~1238-0836, 
 ~1418-0306, ~1428-1950,   ~1439-1338, ~2222-2337.~~ }
The spectra of these objects show characteristics of C-R stars. 
 The spectra of HE~1011-0942,  HE~1205-0521, HE~1238-0836 match closest to 
the spectrum of RV~Sct. The effective temperatures of the objects estimated
using (J-K) calibration range from 3521 K (HE~1238-0836) to 4875 K
(HE~1127-0604).

In the spectra of HE~1238-0836 and HE~1428-1950 the CH band  around 4300 \AA\, 
is slightly  deeper than in RV~Sct. The molecular features in the redward of 
5700 \AA\, appear marginally weaker. 
In the spectra of HE~0945-0813, the  CN band around 4215 \AA\, is 
distinctly detected.  Ca I line at 4226 \AA\, which is generally very
weak or absent in CH stars appears very strongly in the spectrum
of this  object.  The feature due to Na I D   appears very strong.
Feature due to the secondary P-branch head around 4342 \AA\,  is 
 somewhat noticeable in  the spectrum of  HE~0945-0813. The lines due to 
Ba II at 6496 \AA\, and
H$_{\alpha}$ are detected in both the spectra. The molecular bands around 4733, 
5165 and 5635 \AA\,  appear  strongly   in the spectrum of 
HE~0945-0813.  
The spectrum of HE~2222-2337  has low flux below about 4100 \AA\,.
The CH band does not appear as strong as it should be in C-R star's spectrum.
The  CN  band around 4215 \AA\, is marginally 
detected. Other molecular  bands are of similar 
strengths. The  molecular features redward of  5635  \AA\, are slightly weaker. 
We place these objects in  the  C-R group.\\

{\bf  HE~0037-0654,  ~0954+0137,  ~1230-0327, ~1400-1113, ~ 1430-0919, 
~1447-0102, ~2157-2125, ~2216-0202,  ~2255-1724, ~2305-1427, ~2334-1723,
 ~2347-0658, ~2353-2314.~~}
The spectra of these objects are characterized by a  weak (or absent) CN band 
around 4215 \AA\,. Apart from this feature the spectra are somewhat similar
 to the  spectrum of HD~209621.

The spectrum of HE~1230-0327 shows a strong  G-band of CH
and  a  distinct  secondary P-branch head near 4342 \AA\,. Ca I feature
at 4226 \AA\, is not detected. The CN band around 4215 \AA\, is almost 
absent. While
atomic lines of  Ca II K, H , H$_{\alpha}$, Na  I D are distinctly seen,
 BaII line  at 6496 \AA\,  is marginally detected.
The spectra of   HE~0954+0137,  HE~1400-1113, 
HE ~1430-0919, HE~2255-1724 and  HE~2347-0658 look very similar to the spectrum
 of HE~1230-0327.
In the spectra  of these objects  the feature  due to the CN band 
around 4215 \AA\
is marginally detectable. Weak molecular bands noticed in the 
spectrum of HD~209621 upward of 5700 \AA\, are  not observable in 
these spectra. Compared to 
 HD~209621, the molecular bands around 4733, 5165, and 
5635 \AA\, are slightly weaker in the spectra of these  objects. 
Ca II K and H are detected almost with equal strength as in HD~209621. 
 In the spectrum of HE~1430-0919 the  secondary P-branch head near  4342 \AA\,
 is  marginally weaker than  in HD~209621. While the molecular band  
around 5165 \AA\ shows a good match, the 
bands around 4733 and 5635 \AA\, are marginally stronger. The spectrum 
in the redward of 5700 \AA\, resembles  the spectrum of HD~209621.
Features due to Na I D, H$_{\alpha}$ and Ba II line at  6496 \AA\, are detected.
In the spectrum of HE~1447+0102, the CN  band around 4215 \AA\, is almost
absent. Strong well defined  features  of Ca II K and H  are seen. 
The C$_{2}$ molecular bands around 4733, 5165, 5635 \AA\, 
 are distinctly present. No other molecular  bands are noticed longward of 
5700 \AA\,.
The spectra of HE~2305-1427 and  HE~2334-1723   show the  CN band around
 4215 \AA\, with band depth almost half of that in HD~209621. 
All other molecular features  match well  with their counterparts in the
spectrum of  HD~209621. Weak
 molecular bands that  are noticed in the spectrum of HD~209621 upward 
of 5700 \AA\,
are not noticeable in the spectra  of these two  objects. The features due to 
Na  I D, H$_{\alpha}$ and Ba II at 6496 \AA\, could be detected.
The secondary P-branch head at 4223 \AA\,  is seen as distinctly as 
in HD~209621.

In the spectrum of HE~2353-2314, the CH band around 4300 \AA\, as well as the
 CN  band  around 4215 \AA\,
 are  marginally detected. The carbon molecular band around  5165 \AA\,  is 
clearly detected; the band around 5635 \AA\, is much  weaker. No other 
molecular  bands or atomic lines are detectable.  
In the spectrum of  HE~2255-1724,  the CN band  around 4215 \AA\ is much 
weaker   
 than that in HD~209621. The CH band and Ca II K and H are of
similar depths. The molecular bands around  4733, 5165, 5635 \AA\, are slightly 
weaker in this object.  Features of  Na I D,  Ba II line at 6496 \AA\,, and
 H$_{\alpha}$  are distinctly seen. 
 The molecular  bands longward of 5635 \AA\, are not detectable. 
The spectrum acquired on  Sep 11, 2008 have a better signal. In the
 spectrum of HE~2334-1723, the CN band around 4215 \AA\,  is much weaker than 
in HD~209621; all other molecular  bands show a good match.
 The features due to Ca II K  and H also  show a good match.  No molecular  
bands  are  detectable upward of 5700 \AA\,.
In the spectrum of HE~2347-0658 the CN band  around 4215 \AA\, is almost 
absent. 
Ca II K and H features and  carbon molecular bands around 4733, 5165, 
5635 \AA\
show a good match. Molecular  bands seen in HD~209621 upward of 5700 \AA\,
 are not detectable in the spectrum of this object.

 The spectrum of HE~1400-1113 is noisy  below about 4220 \AA\,. The CN band 
around 4215 \AA\,
could be marginally detected. Ca  II K and H  are detected as  weak features.
 A  strong  CH band around 4300 \AA\,  and the secondary  P-branch head 
near 4342 \AA\,  are distinctly seen. Other molecular
 features have band depths 
marginally weaker than their counterparts in HD~209621. Except  for 
Na  I D, Ba II at  6496 \AA\, and 
H$_{\alpha}$ no other atomic lines are detected redward of 5670 \AA\,. 

The spectrum of HE~1430-0919  also shows a very weak CN band around 4215 \AA\,. 
The features
due to Ca II K and H are not detected. The G-band of CH around 4300 \AA\, is 
however very strong in the spectrum.
The spectrum of HE~2157-2125 shows  the CH band around 4300 \AA\, with almost 
equal 
strength to its counterpart  in HD~209621. Features due to  Ca II K  and H 
 and other 
molecular features are  also seen with equal intensities.  However,  
the CN band around 4215 \AA\,  is much 
weaker than that in HD~209621. The spectrum obtained in October, 2008 has a
better signal  than the spectra  obtained in  June and September, 2008.

 The spectrum of HE~0037-0654  looks very similar to the spectrum
of HD~26; however, molecular  bands  of C$_{2}$ around 4730, 5165 and 
5635 \AA\, are marginally stronger than their counterparts in  HD~26. The
CN band around 4215 \AA\, is barely detected, much  weaker than in HD~26.
 Ba II line at 6496 \AA\, is clearly detected. Strong lines 
of H$_{\alpha}$ and Na  I D are   distinctly noticed. 
Except for the features of  Ca II K  and H  which are much weaker,
the spectrum of HE~2216-0202 is  very similar to  the spectrum of  HD~26.
 The secondary 
P-branch head around 4342 \AA\,  is  much stronger than in HD~26. 
The CN band around 4215 \AA\, is not
observed. The CH band at 4305 \AA\, is not  as strong as in HD~26. The molecular
bands around  4733 \AA\, and   5236 \AA\, are  of similar depths. The carbon 
molecular band around 
5635 \AA\, is weaker than the band around 5165 \AA\,. No molecular bands 
longward of 5700 \AA\, are detectable. 

The spectrum of HE~1315-2035 is  noisy blueward 
of 4200 \AA\,.  The G-band of CH and the  carbon  molecular  bands near 
 4733, 5165 and  5635 \AA\,  are   detected in the spectrum. The H$_{\alpha}$  
feature  is  clearly  detected. The effective temperature of this object is 
4639 K as estimated  from  (J-K) colour calibration.

{\bf HE~1027-2644.~~} The spectra of HE~1027-2644 do not show presence 
of  any carbon molecular bands. The features due to  Ca  II K and H are 
 not detected. The  G-band of CH is seen as a weak feature. 
Features due to Ca I at 4226 \AA\, and Na D I are 
  seen as strong features.  Ba II line at 6496 \AA\, and H$_{\alpha}$ 
feature are detected. 2MASS JHK photometry is not available for this object.

\begin{table}
\centering
{\bf Table 4: Estimated effective temperatures (T$_{eff}$) from 
semi-empirical relations}

\begin{tabular}{|c|c|c|}
\hline\noalign{\smallskip}
Star Name &    Teff(J-K)  &  Teff(J-H)  \\
\noalign{\smallskip}\hline\noalign{\smallskip}
HE0008-1712 &  4556.52& 4378.94(-0.5) \\
            &         & 4446.43(-2.5)\\ 
            &         &                            \\
HE0009-1824 &  5530.27& 5377.40(-0.5) \\ 
            &         & 5443.08(-2.5) \\
            &         &                            \\
HE0037-0654 &  5496.71& 4912.74(-0.5) \\ 
            &         & 4980.33(-2.5) \\
            &         &                            \\
HE0052-0543 &  3924.82 &3831.70(-0.5) \\ 
            &          &3896.64(-2.5) \\
            &         &               \\
HE0100-1619 &  4615.21 &4305.94(-0.5) \\
            &          &4373.23(-2.5) \\
            &         &                 \\
HE0113+0110 &  4130.72 &3973.94(-0.5)  \\
            &         & 4039.78(-2.5) \\
            &         &                  \\
HE0136-1831&   4459.39& 4493.10(-0.5)  \\
           &          & 4560.81(-2.5) \\
            &         &                \\
HE0217+0056&   2794.61& 3053.31(-0.5) \\
           &         &  3110.53(-2.5) \\
            &         &                 \\
HE0225-0546&   4051.67& 4086.82(-0.5) \\ 
            &         & 4153.25(-2.5) \\
            &         &                \\
HE0228-0256 &  3655.89& 3916.80(-0.5) \\ 
            &         &  3982.30(-2.5) \\
            &         &                \\
HE0308-1612 &  4420.27& 4428.17(-0.5)  \\
           &          & 4495.77(-2.5)  \\
            &         &                     \\
HE0341-0314&   4119.26& 4336.67(-0.5) \\
           &          & 4404.05(-2.5) \\
            &         &                 \\
HE0420-1037&   4587.42& 4534.65(-0.5) \\ 
           &         &  4602.42(-2.5) \\
            &         &                \\
HE0945-0813 &  4650.36& 4629.30(-0.5)  \\
            &         & 4697.14 (-2.5) \\
            &         &               \\
HE1011-0942 & 3417.52& 3646.15(-0.5) \\
            &         & 3709.67(-2.5) \\
            &         &                \\
HE1019-1136 &  2690.13 &3140.84(-0.5) \\
            &         & 3199.16(-2.5)\\ 
            &         &                \\
HE1028-2501 & 3824.73 & 3769.05(-0.5)  \\ 
            &         & 3833.54(-2.5)  \\
            &         &                \\
HE1045-1434   &4436.49& 4738.21(-0.5) \\ 
               &      & 4806.03(-2.5) \\
            &         &                \\
HE1051-0112  & 4594.34& 4411.69(-0.5) \\
             &        & 4479.26(-2.5) \\
            &         &               \\
HE1102-2142  &4492.46& 4383.27(-0.5) \\
             &       & 4450.77(-2.5)\\
            &         &              \\
\noalign{\smallskip}\hline 
\end{tabular}

\end{table}

\begin{table}
\centering
{\bf Table 4: Estimated effective temperatures (T$_{eff}$) from 
semi-empirical relations (continued)}

\begin{tabular}{|c|c|c|}
\hline\noalign{\smallskip}
Star Name &    Teff(J-K)  &  Teff(J-H)  \\
\noalign{\smallskip}\hline\noalign{\smallskip}
HE1110-0153 &  3969.89& 3842.74(-0.5)  \\
            &         & 3907.75(-2.5) \\
            &         &                  \\
HE1116-1628 &  4224.43 &4125.85(-0.5) \\ 
            &         & 4192.46(-2.5)\\
            &         &               \\
HE1119-1933  & 4675.25& 4790.28(-0.5)\\ 
            &         & 4858.06(-2.5)  \\
            &         &               \\
HE1120-2122 &  4148.01& 3961.57(-0.5)\\
            &         & 4027.33 (-2.5)\\
            &         &                    \\
HE1123-2031 &  4365.87& 4339.85(-0.5)   \\
           &         &  4407.24(-2.5) \\
            &         &                \\
HE1127-0604 &  4875.50& 4604.60(-0.5) \\
             &       &  4672.42(-2.5)  \\
            &         &                  \\
HE1142-2601 & 4475.87 & 4464.64(-0.5)   \\ 
            &         & 4532.31(-2.5)   \\
            &         &                  \\
HE1145-1319   &4485.81 &4514.13(-0.5)  \\
             &      &   4581.87(-2.5) \\
            &         &                \\
HE1146-0151  & 4515.88& 4525.81(-0.5)  \\ 
             &       &  4593.57(-2.5)  \\
            &         &                \\
HE1157-1434 &  4182.97 &4181.08(-0.5) \\
           &         &  4247.93(-2.5)\\
            &         &                \\
HE1205-0521 &  4650.36 &4927.12(-0.5)  \\
            &        &  4994.67(-2.5)\\
            &         &                \\
HE1205-2539&  4182.97 & 4046.42(-0.5)   \\
            &         &  4112.64(-2.5)  \\
            &         &                    \\
HE1228-0417 &  4795.85& 4964.09(-0.5) \\ 
           &          & 5031.55(-2.5)\\
            &         &               \\
HE1230-0327  & 4814.61& 4846.74(-0.5)  \\
               &      & 4914.44(-2.5) \\
            &         &               \\
HE1238-0836&  3521.23 &  3954.13(-0.5)  \\
           &          &  4019.85(-2.5)  \\
            &         &                  \\
HE1253-1859 &  4239.41& 4105.00(-0.5)  \\
            &         & 4171.51(-2.5) \\
            &         &                \\
HE1315-2035 & 4639.76 &4949.26(-0.5) \\
           &          & 5016.77(-2.5)  \\
            &         &                 \\
HE1318-1657 &  4423.50& 4612.48(-0.5)\\
            &         & 4680.31(-2.5)  \\
            &         &                   \\
HE1331-2558  & 4227.41& 4218.84(-0.5) \\
           &           &4285.84(-2.5)  \\
            &         &                  \\
HE1344-0411  & 3118.34 &3414.55(-0.5) \\
            &          &3475.93(-2.5) \\ 
            &         &                \\
\noalign{\smallskip}\hline 
\end{tabular}

\end{table}

\begin{table}
\centering
{\bf Table 4: Estimated effective temperatures (T$_{eff}$) from 
semi-empirical relations (continued)}

\begin{tabular}{|c|c|c|}
\hline\noalign{\smallskip}
Star Name &    Teff(J-K)  &  Teff(J-H)  \\
\noalign{\smallskip}\hline\noalign{\smallskip}
HE1358-2508&  3623.67&  3826.04(-0.5)   \\
            &         &  3890.93(-2.5)  \\
            &         &                 \\
HE1400-1113 &  4894.81& 5010.94(-0.5) \\
            &         & 5078.27(-2.5) \\
            &         &                \\
HE1404-0846 &  4239.40& 4027.25(-0.5)\\
            &         & 4093.38(-2.5) \\
            &         &                \\
HE1405-0346  & 4391.32& 4484.40(-0.5) \\
            &        &  4552.09(-2.5) \\
            &         &                 \\
HE1410-0125&  4378.56  &4266.53(-0.5) \\
           &          & 4333.70(-2.5) \\
            &         &              \\
HE1418-0306 & 3598.71&  3761.98(-0.5) \\
            &         & 3826.41(-2.5) \\
            &         &                \\
HE1425-2052&  4191.80 & 4250.49(-0.5)  \\
           &          & 4317.60 (-2.5)  \\
            &         &                  \\
HE1428-1950 & 4505.82  &4573.18 (-0.5)\\
            &         & 4640.98(-2.5) \\
            &         &                \\
HE1429-1411 & 3057.76 & 3302.55(-0.5)  \\
            &        &  3362.74(-2.5)  \\
            &         &                \\
HE1430-0919 & 4700.38&  4625.80(-0.5) \\
            &         & 4693.64(-2.5)\\
            &         &                \\
HE1431-0755 & 3937.98 & 3950.23(-0.5)\\
             &        & 4015.92(-2.5)  \\
            &         &                \\
HE1432-2138 & 5074.94 & 5075.25(-0.5)\\
            &         & 5142.38(-2.5) \\
            &         &                     \\
HE1439-1338 &3658.21  & 3898.38(-0.5)\\
            &         & 3963.76(-2.5) \\
            &         &                   \\
HE1440-1511 & 4636.24  &4747.85(-0.5)\\
            &         & 4815.66 (-2.5)\\
            &         &               \\
HE1442-0346 & 4622.20 & 4655.40(-0.5) \\
            &         & 4723.24(-2.5)\\
            &         &                \\
HE1447+0102 & 5042.06 & 4893.55 (-0.5)\\
            &         & 4961.18(-2.5) \\
            &         &                \\
HE1525-0516 & 4546.30 &4695.10 (-0.5)\\
            &         & 4762.94(-2.5) \\
            &         &                 \\
HE2114-0603 & 3948.56 & 3919.97(-0.5)  \\
            &         & 3985.48(-2.5) \\
            &         &                \\
HE2157-2125 & 4583.97 & 5135.81(-0.5)\\
            &        &  5202.71(-2.5)\\ 
            &         &                \\
HE2211-0605 & 4553.10 &4621.90(-0.5) \\ 
            &         & 4689.73(-2.5) \\ 
            &         &                \\
\noalign{\smallskip}\hline 
\end{tabular}
\end{table}

{\footnotesize
\begin{table}
\centering
{\bf Table 4: Estimated effective temperatures (T$_{eff}$) from 
semi-empirical relations (continued)}

\begin{tabular}{|c|c|c|}
\hline\noalign{\smallskip}
Star Name &    Teff(J-K)  &  Teff(J-H)  \\
\noalign{\smallskip}\hline\noalign{\smallskip}
HE2213-0017 & 2701.10 & 3005.84(-0.5)\\ 
            &         & 3062.45(-2.5) \\ 
            &         &                  \\
HE2216-0202 & 4969.45&  5122.15(-0.5) \\ 
            &         & 5189.11(-2.5) \\
            &         &                  \\
HE2222-2337 & 3911.73 & 3879.01(-0.5) \\ 
            &         & 3944.26(-2.5) \\ 
            &         &                 \\
HE2225-1401 & 2169.18 & 2845.70(-0.5)\\ 
             &        & 2900.11(-2.5) \\ 
            &         &                \\
HE2228-0137 & 4369.03 & 4284.06(-0.5) \\ 
            &        &  4351.28(-2.5) \\ 
            &         &                \\
HE2246-1312 & 4110.70&  4125.95(-0.5) \\ 
            &        &  4192.57(-2.5)\\ 
            &         &                   \\
HE2255-1724 & 4675.26&  4388.73(-0.5)\\ 
            &        &  4456.25(-2.5) \\ 
            &         &                  \\
HE2305-1427 & 5042.06&  4594.23(-0.5) \\ 
            &        &  4662.05(-2.5) \\ 
            &         &                \\
HE2334-1723 &4729.39 &4827.13 (-0.5) \\ 
             &        & 4894.87(-2.5) \\ 
            &         &                  \\
HE2347-0658 & 5554.46 &4989.70(-0.5)  \\ 
            &         & 5057.10(-2.5) \\ 
            &         &               \\
HE2353-2314 & 3927.44 &4014.92(-0.5)\\ 
            &         & 4080.98(-2.5)\\ 
\noalign{\smallskip}\hline 
\end{tabular}

The numbers inside the parentheses indicate the adopted metallicities [Fe/H] \\
\end{table}
}

{\footnotesize
\begin{table}
\centering
{\bf Table 5: Stars with   radial velocity estimates  }
\begin{tabular}{|c|c|c|}
\hline\noalign{\smallskip}
Star Name &   $v_{\rm r}$ km s$^{-1}$  &  Reference  \\
\noalign{\smallskip}\hline\noalign{\smallskip}
HE~0100$-$1629 & $-$142.0              & 1   \\
HE~0217$+$0056 & $-$142.0 ${\pm}$ 3    & 2  \\
HE~0228$-$0256 & $-$72.0               & 1   \\
HE~1019$-$1136 & $-$126.0 ${\pm}$ 4    & 2  \\
HE~1105$-$2736 & $-$36.0 ${\pm}$ 1.3    & 3  \\
HE~1116$-$1628 & $-$69.0               & 2 \\
HE~1152$-$0355 & $+$431.3 ${\pm}$ 1.5 & 4  \\
HE~1305$+$0007 & $+$217.8 ${\pm}$ 1.5 & 4 \\
HE~1410$-$0125 & $+$88.0 ${\pm}$ 3     & 5    \\
HE~1429$-$1411 & $-$90.0 ${\pm}$ 1.5   & 6  \\
HE~1442$-$0058 & $-$37.0 ${\pm}$ 4     & 2   \\
HE~2213$-$0017 & $-$44.0 ${\pm}$ 3    & 2 \\
HE~2225$-$1401 & $-$113.0 ${\pm}$ 5    & 2  \\
HD~26        & $+$217.8 ${\pm}$ 1.5 & 6  \\
HD~5223      & $-$244.9 ${\pm}$ 1.5 & 4  \\
HD~209621    &  $-$390.5 ${\pm}$ 1.5 & 6  \\
\noalign{\smallskip}\hline 
\end{tabular}

References: 1: Bothun et al. (1991), 2: Totten \& Irwin (1998), 
3: Zwitter et al. (2008), 4: Goswami et al. (2006), 5: Frebel et al. (2006), 
6: Goswami et al. (in preparation)
\end{table}
}

\section{Concluding Remarks}

An accurate assessment of the  fraction of CH stars can significantly aid 
our understanding of formation and evolution of heavy elements at low 
metallicity. Another important issue is the role of low to intermediate-mass
stars of the halo in the early Galactic chemical evolution. Thus large samples
of faint high latitude  stars such as the one reported by Christlieb et al. 
(2001b)  that contain different types of carbon stars need to be analyzed to 
understand the astrophysical implications  of each individual type  of 
stellar population.  Our objective  in this study has been to identify the 
CH stars (as well as different type of stellar objects) in a selected sample 
of high  Galactic latitude field stars. The sample is  based on  our  
on-going  observational programs with HCT and VBT  on cool  stars. During 
2007 and 2009 we have acquired low resolution spectra for a large number of 
stars that included  about ninety two  objects from the Hamburg survey of 
Chrisltieb et al. (2001b).

The  spectral classification criteria are those  presented in Goswami (2005).
Among the ninety two objects, the spectra of  seventy  objects are 
characterized by the presence of strong C$_{2}$ molecular bands. The spectra
of  twenty two  objects show only a weak or moderate  G-band of CH and a CN 
band 
around 4215 \AA\,.  One object, HE~1027-2644 does not show presence of any 
molecular bands in its spectrum. The spectral   analysis  led to the 
detection of 
thirty six  potential CH star candidates.  Their locations on the two
color J-H versus H-K diagram, estimated effective temperatures, and carbon
isotopic ratios are in support of their classification with this class
of objects. This set of objects  will make important targets for  subsequent
chemical composition  studies  based on high resolution spectroscopy and for 
confirmation of these objects with this class of identification.

 While identification of C-N and C-J type stars are relatively easy,  
separating C-R stars from CH stars is not so straightforward.
The two main properties, presence or absence of $s-$process elements and
binarity that differentiate early-R stars from CH stars, can
be known  only through detailed abundance studies that require high
resolution spectroscopy and from long-term radial velocity monitoring.
 The faintness of these objects makes   high resolution spectroscopic
studies  arduous and time-consuming. As such, the method described in
Goswami (2005) to distinguish a C-R from a CH star proved  quite
useful (Goswami et al. 2006).
High resolution spectra will also allow for    an accurate 
measurement of $^{12}$C/$^{13}$C ratios.  Based on medium resolution spectra,
 for  the potential CH star candidates,  we find a low (${\le}$ 10)  
carbon isotopic ratios,  indicating that they belong 
to  the group of early-type CH stars.

Abia et al. (2002) have shown that CH stars  cannot be formed above a 
threshold metallicity, around Z $\sim$ 0.4Z$_{\odot}$.  According to 
Dominy (1984), the metallicities of C-R stars  are either  solar or slightly 
sub-solar. C-R stars are believed to be Core Helium Burning (CHeB) 
counterparts of CH stars in which $s-$process elements are either absent 
or not detectable (Izzard et al. 2007).  These authors  have predicted an 
early-R/CH ratio $\sim$ 7\%, at [Fe/H] = -2.3, a metallicity typical of the
Galactic halo.  This ratio  is derived considering only CHeB CH stars,  if 
CH giants and dwarfs are also considered, this ratio is likely to get much
lower.

Westerlund et al. (1995) defined dwarf carbon stars as having J-H ${\le}$ 0.75,
 H-K ${\ge}$ 0.25 mag. Among the three dwarf carbon stars in our sample the
 objects HE~1358-2508 occupies a region defined by these limits on J-H, H-K 
plane.  HE~0009-1824  and HE~1116-1628 however do not follow the JHK 
definition of dwarf carbon stars offered by Westerlund et al. (1995). It 
seems, these limits on J-H and H-K may  not be  very tight. Proper motions
of these objects have been estimated by Mauron et al. (2007) and have 
placed them as  dwarf carbon stars.

 The temperature estimates of the program stars derived using JHK-temperature
 calibration relations of Alonso et al. (1996), although varying over a wide 
range,  provide a  preliminary temperature check for the program stars and 
can  be 
used  as starting values in deriving atmospheric  parameters from high 
resolution spectra using model atmospheres. 

Important information such as kinematic properties of the stars can be 
derived from radial velocity estimates.
From low resolution spectra, radial velocities are generally computed using
 Fourier cross-correlation method. This method,
widely  employed, uses  the spectrum of a radial velocity standard as the 
template spectrum.  Unfortunately, we could  not acquire spectra of radial 
velocity standards that are usable for the present set of program stars 
under study. 
Estimated radial velocities using a few atomic lines,
detectable on the low resolution spectra,  did not return consistent results.
However,  a systematic estimation of radial velocities
of the program stars using appropriate radial velocity standard templates
would be a worthwhile future program.
Radial velocities of a few stars belonging to our program star list,
that are available in literature, are listed in Table 5.

The primary focus of  this work is CH stars; however, a  detailed discussion on
the objects of other  spectral types is also necessary and is under progress.\\

 {\it Acknowledgement}\\
 We thank  the staff at IAO, VBO and at the remote control station  at CREST,
Hosakote for assistance during the observations.  This work made use of the
SIMBAD astronomical database, operated at CDS, Strasbourg, France, and the
NASA ADS, USA. Ms Drisya K. is a JRF in the DST project NO. SR/S2/HEP-09/2007; 
funding from the above mentioned project is greatfully acknowledged. 
\\


\begin{thebibliography}{}
\bibitem {}  Abia C., Dominguez I., Gallino R., Busso M., Masera S., Straniero
O., de Laverny P., Plez B., Isern J., 2002, ApJ, 579, 817
\bibitem {}  Alonso A., Arribas S. \& Martinez-Roger C., 1994, A\&AS, 107, 365
\bibitem {}  Alonso A., Arribas S. \& Martinez-Roger C., 1996, A\&A, 313, 873
\bibitem {}  Alonso A., Arribas S. \& Martinez-Roger C., 1998, A\&AS, 131, 209
 \bibitem {}  Aoki, W., Norris, J. E., Ryan, S. G., Beers, T. C. \& Ando, H.,
2002, ApJ, 567, 1166
\bibitem {}   Aoki  W., Ryan  S. G., Norris  J. E., Beers  T. C., Ando  H., \&
Tsangarides  S., 2002, ApJ, 580, 1149
\bibitem{}  Aoki, W., Beers, T.C., Christlieb, N., Norris, J.E., Ryan, S.G., \&
Tsangarides, S., 2007, ApJ, 655, 492
\bibitem {}  Barnbaum, C., Stone, R. P. S. \& Keenan, P., 1996, ApJS, 105, 419
\bibitem {} Bonifacio P., Molaro P., Beers T. C., Vladilo G., 1998, A\&A, 332, 672
\bibitem {}  Bothun G., Elias J. H., MacAlpine G., Mathews K., Mould J. R., 
Neugebauer G., Reid I. N., 1991, AJ, 101, No.6, 2220
\bibitem {} Christlieb N., Wisotzki L., Reimers D., Homeier D., Koester D., \&
Heber, U., 2001a, A\&A, 366, 898
\bibitem {}  Christlieb N., Green P. J., Wisotzki L., Reimers D., 2001b,
A\&A, 375, 366
\bibitem {} Cohen, J. G., Shectman S., Thompson I., McWillium A., Christlieb
 N., Melendez J., Zickgraf Franz-josef, Ramirez S., Swenson A., 2005, ApJ, 633, L109
\bibitem {}  Dominy, J. F., 1984, ApJS, 55, 27
\bibitem {} Frebel, A., Chriestlieb, N., Norris John E., Beers Timothy C.,
Bessell, Michael S., Rhee Jaehon, Fechner Cora, Marsteller Brian, et al., 2006,
ApJ, 652, 1585
\bibitem {} Goswami, A., Bama P., Shantikumar, N. S., Devassy Deepthi, 2007,
MNRAS, 35, 339 
\bibitem {} Goswami A., 2005, MNRAS, 359, 531
\bibitem {}  Goswami, A., Aoki W., Beers T. C., Chriestlieb N., Norris J. E.,
 Ryan S. G., Tsangarides S., 2006, MNRAS, 372, 343
\bibitem {}  Green  P. J., Margon B., Anderson S. F., Cook K. H., 1994,
ApJ, 434, 319
\bibitem {}  Green  P. J. \& Margon, B., 1994,  ApJ, 423, 723
\bibitem {}  Green, P. J., Margon, B., Anderson  S. F., MacConnell D. J., 1992,
 ApJ, 400, 659
\bibitem {}  Hartwick, F. D. A. \& Cowley, A. P. 1985, AJ, 90, 2244
\bibitem {} Hill V.,  Barbuy B., Spite M., Spite F., Cayrel R., Plez B., Beers T. C., Nordstrom B., Nissen P. E.,  2000, A\&A, 353, 557
\bibitem {} Izzard R. G., Jeffery C. S., Lattanzio J., 2007, A\&A, 470, 661 
\bibitem {}  Keenan Philip C. 1993, PASP, 105, 905
\bibitem {}  Lambert, D. L., Gustafsson, B., Eriksson, K., Hinkle, K. H., 1986, ApJS, 62, 373
\bibitem {}Lowrance Patrick J., Kirkpatrick J. Davy, Reid I. Neill, Cruz Kelle L., Liebert James, 2003, ApJ, 584, L95 
\bibitem {}  Lucatello S., Gratton R. G., Beers T. C., Carretta E.,  2005,
ApJ 625, L833-837
\bibitem {} Mauron, N., Gigoyan K. S., Kendall T. R., 2007,  A\&A, 463, 969 
\bibitem {}  McClure, R. D. \& Woodsworth, A. W., 1990, ApJ, 352, 709
\bibitem {}  McClure, R. D., 1983, ApJ, 268, 264
\bibitem {}  McClure, R. D., 1984, ApJ, 280, L31
\bibitem {}  Norris, J. E., Ryan, S. G., \& Beers, T. C., 1997a, ApJ, 488, 350
\bibitem {}  Norris, J. E., Ryan, S. G., \& Beers, T. C., 1997b, ApJ, 489, L169
\bibitem {}  Norris, J. E., Ryan, S. G., Beers, T. C., Aoki, W. \&  Ando, H.,
2002, ApJ, 569, L107
\bibitem {}  Ramirez  I.,   Melendez, J., 2004, ApJ, 609, 417
\bibitem {} Ratnatunga, K. U., 1983, PhD Thesis, Australian National Observatory
\bibitem {} Rossi, S., Beers, T. C. \& Sneden, C. 1999, in ASP Conf Ser. 165,
{\it The Third Stromlo Symposium: the Galactic Halo, ed. B. K. Gibson, T. S.
Axelrod \& M. E. Putman } (San Francisco: ASP), 264
\bibitem {}  Totten, E. J., Irwin M. J., Whitelock P. A., 2000, MNRAS, 314, 630
\bibitem {}  Totten, E. J. \& Irwin, M. J. 1998, MNRAS, 294, 1
\bibitem {}  Tsuji, T., Tomioka K., Sato H., Iye M., Okada T.,  et al.,
 1991, A\&A, 252, L1
\bibitem {}  Vanture, Andrew D., 1992, AJ, 104, 1997
\bibitem {}  Wallerstein, G. \& Knapp, G. 1998, ARA\&A, 36, 369 
\bibitem {}  Westerlund, B. E., Azzopardi, M., Breysacher, J. 
Rebeirot, E., 1995, A\&A, 303, 107
\bibitem {} Wisotzki, L., Christlieb N., Bade, N., Beckmann, V., Kohler T., 
Vanelle, C., Reimers, D., 2000, A\&A, 358, 77
\bibitem {}  Yamashita, Y., 1975, PASJ, 27, 325
\bibitem {} Zijlstra, A. A.,  2004, MNRAS, 348, L23
\bibitem {}  Zinn, R., 1985, ApJ, 293, 424
\bibitem {} Zwitter, T., Siebert A., Munari U., Freeman K. C., Siviero, A.,
Watson, F. G., Fullbright, J. P., Wyse, R. F. G. et al., 2008, AJ, 136, 421

\end{thebibliography}
\end{document}